\documentclass[reprint,aps,prl,amsmath,amssymb,superscriptaddress, preprintnumbers]{revtex4-2}
 
\usepackage[utf8]{inputenc}
\usepackage[T1]{fontenc}

\usepackage{amsmath}
\usepackage{mathrsfs}
\usepackage{txfonts}
\usepackage{mathtools}
\usepackage{braket}
\usepackage{tensor}
\usepackage{xcolor}
\usepackage[abbreviations]{siunitx}
\usepackage{graphicx}
\usepackage{booktabs}

\usepackage{float}

\usepackage{braket}
\usepackage{ulem}

\usepackage{hyperref}
\usepackage{cleveref}

\newcommand{\beq}{\begin{equation}}
\newcommand{\eeq}{\end{equation}}
\newcommand{\beqn}{\begin{eqnarray}}
\newcommand{\eeqn}{\end{eqnarray}}
\newcommand{\bsub}{\begin{subequations}}
\newcommand{\esub}{\end{subequations}}
\newcommand{\bpm}{\begin{pmatrix}}
\newcommand{\epm}{\end{pmatrix}}

\usepackage[dvipsnames]{xcolor}

\begin{document}

\title{Quantum effects in the quadrupole rotor picture of ultra-relativistic ion-ion collisions}

\author{Stavros Bofos}
\affiliation{CEA, DES, IRESNE, DER, SPRC, LEPh, 13108 Saint-Paul-lez-Durance, France}

\author{Yi Li}
  \affiliation{School of Physics and Astronomy, Sun Yat-sen University, Zhuhai 519082, P.R. China}    
   \affiliation{Guangdong Provincial Key Laboratory of Quantum Metrology and Sensing, Sun Yat-Sen University, Zhuhai 519082, China }

\author{Chenrong Ding}
  \affiliation{School of Physics and Astronomy, Sun Yat-sen University, Zhuhai 519082, P.R. China}    
   \affiliation{Guangdong Provincial Key Laboratory of Quantum Metrology and Sensing, Sun Yat-Sen University, Zhuhai 519082, China } 
   
\author{Benjamin Bally}
\affiliation{Technische Universit\"at Darmstadt, Department of Physics, 64289 Darmstadt, Germany}

\author{Thomas Duguet}
\email{thomas.duguet@cea.fr}
\affiliation{Université Paris-Saclay – CEA – IRFU, 91191 Gif-sur-Yvette, France}  

\author{Mikael Frosini}
\affiliation{CEA, DES, IRESNE, DER, SPRC, LEPh, 13108 Saint-Paul-lez-Durance, France}

\author{Jiangming Yao}
\email{yaojm8@sysu.edu.cn}
  \affiliation{School of Physics and Astronomy, Sun Yat-sen University, Zhuhai 519082, P.R. China}    
   \affiliation{Guangdong Provincial Key Laboratory of Quantum Metrology and Sensing, Sun Yat-Sen University, Zhuhai 519082, China } 
    \affiliation{Yukawa Institute for Theoretical Physics, Kyoto University, Kyoto, 606-8502, Japan}  
    

\begin{abstract}
The azimuthal hadronic flow observed in ultra-relativistic ion-ion collisions provides a sensitive probe of many-body ground-state correlations in the colliding nuclei. In particular, collective correlations associated with nuclear “intrinsic deformation” are expected to leave pronounced fingerprints on specific final-state observables. However, such effects are commonly interpreted within a classical rigid-rotor picture, despite the intrinsically quantum nature of nuclei. In this Letter, the validity of this interpretation is assessed systematically across the nuclear chart by comparing the quantum quadrupole rotor with its classical rigid-rotor limit. Quantum contributions associated with the fermionic nature of the nucleons are shown to be largely independent of shell effects, and hence of the intrinsic deformation. While they account for nearly all of the quantum rotor effective quadrupole deformation in light and/or spherical nuclei, they drop below 10\% in intrinsically well deformed  heavy nuclei. The present letter demonstrates that a quantitative interpretation of nuclear-structure effects in final-state observables requires going beyond the classical rigid-rotor paradigm. Beyond the quantum contributions quantified presently, correlations associated with collective vibrations and with the non-collective nucleonic motion must be further included and characterized.
\end{abstract}

\preprint{}

\maketitle


\paragraph{Introduction.}
The atomic nucleus is a strongly-interacting finite quantum many-body system in which nucleons self organize as a result of both collective and non-collective correlations between them. In particular, collective correlations among nucleons leave characteristic fingerprints in $k$-nucleon ($k>1$) correlation functions which are often interpreted phenomenologically in terms of intrinsic nuclear deformation,  giving rise to the classical pictures of nuclear shape rotations and vibrations, and possibly to clustering phenomena.

In recent years, ultra-central ultra-relativistic ion-ion collisions have been understood to provide an unexpected mean to map out many-body correlations in the ground-state of atomic nuclei~\cite{Giacalone2018a,Giacalone2021a,Summerfield2021a,Bally2022a,Zhang2022a,Ryssens2023a,STAR2024a,Giacalone2025a,Giacalone2025b}.  Following the formation and cool down of a quark-gluon plasma (QGP), anisotropies in the azimuthal flow of hadrons detected in the final state of such collisions result, to leading order, from spatial correlations among nucleons in the collided nuclei~\cite{Niemi:2016}. For example, the anisotropy of the azimuthal flow of two hadrons in the final state of a symmetric ultra-central collision, when averaged over events, can be related to moments of the two-nucleon correlation function in the nuclear ground state of the collided nuclei~\cite{Blaizot:2025scr,Duguet2025a}.

Based on such a quantum mechanical formulation of the link between nuclear structure properties in the initial state and observables in the final state~\cite{Duguet2025a}, it is now possible to go beyond the traditional analysis of ion-ion collisions based on the simplistic classical rigid rotor (CRR) nuclear model~\cite{Duguet:2025qxi}. Indeed, collided nuclei are quantum systems and it is legitimate to question the extent to which the (i) classical (ii) rigid (iii) rotor picture holds. Leaving point (iii) to a forthcoming publication~\footnote{The impact of many-body correlations beyond the quantum rotor has been touched upon in Refs.~\protect\cite{Blaizot2025a,Bofos:2026huw} but still demands a dedicated analysis.}, the present work focuses on points (i) and (ii). This is achieved by quantifying the impact of the two approximations leading from the quantum rotor (QR) description to its CRR counterpart. 

The connection between the QR and CRR pictures first relates to an approximation of the quantum mechanical projection at play in the QR and ensuring that the collided nuclei have good angular momentum, e.g., that the ground-state of even-even nuclei with $J=0$ is spherically symmetric in the laboratory frame where the collisions take place. The validity of such an approximation has been wrongly attributed by practitioners to the shorter time scale characterizing ultra-relativistic collisions compared to the one underlying the angular momentum restoration, knowing that there is in fact no time scale associated with the latter~\cite{Dobaczewski:2025rdi}. A recent study underscored the importance of accounting (approximately) for the quantum entanglement associated with the symmetry restoration in the Monte Carlo Glauber modeling of ultra-relativistic collisions~\cite{Ke:2025}, finding a 6\%-8\% reduction in the second-order fireball eccentricity for central Ne--Ne and Ne--Pb collisions. In this Letter, the impact of the approximate angular-momentum restoration on the quantum-mechanical calculation of the effective deformation extracted from ultra-central symmetric collisions is quantified for the first time. 

The connection between the QR and CRR pictures further relies on neglecting quantum contributions associated with the fermionic nature of the nucleons. This second approximation is also assessed here for the first time, thereby clarifying  missing ingredients in the traditional interpretation of ultra-relativistic ion--ion collisions. 

Eventually, quantum contributions to the QR are shown to be independent of shell effects, and hence of intrinsic deformation. They account for nearly all of the QR effective quadrupole deformation in light and/or spherical nuclei, but drop below 10\% in intrinsically deformed nuclei as the nuclear mass increases. 

\paragraph{Mean-square azimuthal flow vs eccentricity.}
The anisotropy of the azimuthal hadronic flow in the transverse plane of ultra-relativistic ion-ion collisions is characterized by the Fourier coefficients $V_{\ell}$ of multipolarity $\ell$. This anisotropic flow is further related, through hydrodynamic evolution of the QGP, to the normalized eccentricities $\epsilon_{\ell}$ of the entropy density $s(\mathbf{r}_{\perp})$ deposited shortly after the reaction~\cite{Sousa:2024msh}
\begin{equation}
   \epsilon_{\ell} \equiv -\frac{\mathcal{E}^{(1)}_{\ell}}{R^{(1)}_{\ell}} \equiv - \frac{\int_{\mathbf{r}_{\perp}} \mathcal{E}^{(1)}_{\ell}(\mathbf{r}_{\perp}) s(\mathbf{r}_{\perp})}{\int_{\mathbf{r}_{\perp}} R^{(1)}_{\ell}(\mathbf{r}_{\perp}) s(\mathbf{r}_{\perp})} \, , \label{eccentricities}
\end{equation}
where $\mathbf{r} = (\mathbf{r}_{\perp},z) = (r_{\perp},\phi,z)$, such that $\mathbf{r}_{\perp}$ denotes the position vector in the transverse plane whereas $\phi$ specifies the corresponding azimuthal angle. In Eq.~(\ref{eccentricities}), the eccentricity and transverse radius are respectively given by
\begin{equation}
 \mathcal{E}^{(1)}_{\ell}(\mathbf{r}_{\perp}) \equiv r^\ell_{\perp} e^{i\ell \phi}  \,\,\, , \,\,\,  R^{(1)}_{\ell}(\mathbf{r}_{\perp}) \equiv r^\ell_{\perp} \, .
\end{equation}

The anisotropy of the azimuthal flow between two particles in the final state is quantified via the variance of $V_{\ell}$ over a large set of events. In ultra-central collisions of present interest, such a variance is known to be, to a good approximation, proportional to the normalized variance $\langle \delta \epsilon^{2}_{\ell}   \rangle$ of the same multipolarity~\cite{Niemi:2016,Sousa:2024msh}. In symmetric collisions, $\langle \delta \epsilon^{2}_{\ell}   \rangle$  was further shown to be related to the quantum mechanical expectation value of the mean-squared eccentricity operator $\mathcal{E}^{(2)}_{\ell}$ in the $J=0$ ground-state $| \Psi^{0^+}\rangle$ of the collided nuclei~\cite{Duguet2025a} according to
\begin{align}
    \langle \delta \epsilon^{2}_{\ell}   \rangle &= \frac{1}{2} \frac{\langle \Psi^{0^+}  | \mathcal{E}^{(2)}_{\ell}| \Psi^{0^+}\rangle }{\langle   \Psi^{0^+}  |  R^{(1)}_{\ell} | \Psi^{0^+}  \rangle^2} \label{meansquareeccentricity}  \\
    &= \frac{1}{2} 
    \frac{\int_{\bold{r}_{1}} \mathcal{E}^{(2) \, (1\text{b})}_{\ell}(\bold{r}_{1}) \rho^{(1)}(\bold{r}_{1}) +\int_{\mathbf{r}_{1,2}}  \mathcal{E}^{(2) \,  \, (2\text{b})}_{\ell}(\bold{r}_{1},\bold{r}_{2}) \rho^{(2)}(\bold{r}_{1}, \bold{r}_{2})}{\Big[ \int_{\bold{r}_{1}} R^{(1)}_{\ell}(\bold{r}_{1}) \rho^{(1)}(\bold{r}_{1}) \Big]^2} \, ,\nonumber
\end{align}
where the one-body $\mathcal{E}^{(2) \, (1\text{b})}_{\ell}$ and two-body $\mathcal{E}^{(2) \, (2\text{b})}_{\ell}$ contributions to the mean-square eccentricity operator $\mathcal{E}^{(2)}_{\ell}$, along with the ground-state local one-body $\rho^{(1)}(\bold{r}_{1})$ and two-body $\rho^{(2)}(\bold{r}_{1},\bold{r}_{2})$ densities, are defined in the supplemental material (SM).

\paragraph{Classical rigid rotor and effective deformation.}
One way to formulate the rigid-rotor (RR) model is to stipulate that the laboratory $k$-body density matrix $\rho^{(k)}_{\text{RR}}$  of a given $J$ state is obtained by averaging the intrinsic $k$-body density matrix $\rho^{(k)}_{\Omega}$ pointing in direction $\Omega\equiv (\alpha,\beta,\gamma)$ over all spatial orientations with appropriate weights given by the Wigner functions ${\cal D}^J_{MK}(\Omega)$. Thus, the RR one- and two-body local densities of a $J=0$ state read as
\begin{subequations}
\label{kbodydensityRR}
\begin{align}
\rho^{(1)}_{\text{RR}}(\mathbf{r}_1)
         &= \frac{1}{8\pi^2} \int_{\Omega}  
       \rho^{(1)}_{\Omega}\big(\mathbf{r}_1\big) \, , \label{kbodydensityRR1}
       \\
\rho^{(2)}_{\text{RR}} (\mathbf{r}_1,\mathbf{r}_2)
        &= \frac{1}{8\pi^2} \int_{\Omega}  
       \rho^{(2)}_{\Omega} (\mathbf{r}_1,\mathbf{r}_2)  \, , \label{kbodydensityRR2}
\end{align}
\end{subequations}
such that the former only depends on $r_1$ and is thus spherically symmetric, whereas the latter  depends on $r_1,r_2$ and the relative angle $\cos\zeta \equiv \mathbf{r}_1 \cdot \mathbf{r}_2/r_1 r_2$. 

The actual {\it classical} RR further relies on the hypothesis that the system displays no correlations {\it in the intrinsic frame} such that the intrinsic $k$-body density is simply the product of $k$ 1-body densities, e.g., 
\begin{align}
\rho^{(2)}_{\Omega} (\mathbf{r}_1,\mathbf{r}_2)
        &= 
       \rho^{(1)}_{\Omega}\big(\mathbf{r}_1\big) \, 
       \rho^{(1)}_{\Omega}\big(\mathbf{r}_2\big)  \, . \label{nocorrintrinsic}
\end{align}
Employing a simple analytical model of the 1-body intrinsic density characterized by the axial quadrupole intrinsic deformation parameter $\beta_{20}$, the normalized quadrupole mean-square eccentricity [Eq.~\eqref{meansquareeccentricity}] of the CRR reads as~\cite{Duguet2025a,Bofos:2026huw}
\begin{equation}
    \langle \delta \epsilon^2_2 \rangle^{\text{CRR}}_{0^+} = \frac{1}{A} + \frac{3}{4\pi} \beta^2_{20} \, , \label{eccCRR}
\end{equation}
i.e., while the trivial one-body contribution is equal to $A^{-1}$ and independent of the intrinsic deformation parameter, the two-body contribution probing genuine two-body correlations is independent of $A$ but scales with the square of the intrinsic deformation parameter. Based on this result, a quantity homogeneous to the square of a dimensionless quadrupole effective deformation parameter is introduced according to~\cite{Blaizot2025a}
\begin{equation}
    \label{eq: Beta parameter definition1}
    \mathcal{B}^2_{2}(\text{HE}) \equiv \frac{4\pi}{3} \langle  \delta \epsilon^{2 \, (2\text{b})}_2 \rangle \; .
\end{equation}
By definition, $\mathcal{B}^2_{2}(\text{HE})$ reduces to the square of the intrinsic quadrupole deformation in the CRR model, i.e., $\mathcal{B}^{2}_{2}(\text{HE})_{\text{CRR}} = \beta^2_{20}$. 

\paragraph{The quantum rotor and its classical rigid approximation.}
The present study focuses on results obtained from the QR description of nuclei provided by projected Hartree-Fock Bogoliubov (PHFB) calculations. This approximation relies on the following ansatz for the $J^\pi=0^+$ ground-state of even-even nuclei
\begin{equation}
| \Theta^{0^+}_{\text{PHFB}} \rangle \equiv \frac{1}{c_{0}} P^{J=0} P^{\text{N}\text{Z}}  | \Phi \rangle \, , \label{PHFBansatz}
\end{equation}
where the normalized symmetry-breaking, i.e., deformed, HFB state $| \Phi \rangle$ displays the {\it intrinsic} axial quadrupole deformation
\begin{align}
\beta_{20}[\text{dHFB}] &\equiv \frac{4\pi}{5} \frac{\langle \Phi | Q_{20} | \Phi \rangle}{\langle  \Phi | r^2 |  \Phi \rangle}\, , 
\end{align}
with the multipole moment operator defined in terms of the nucleons position and spherical harmonics by
\begin{equation}
    \label{eq: quadupole operator definition}
    Q_{\ell \mu} \equiv \sum_{i_1=1}^{A} r^\ell_{i_1} Y^{\mu}_{\ell}(\Omega_{i_1})\; .
\end{equation}
The analytical form of the operator $P^{J}$ ($P^{\text{N}\text{Z}}$) projecting  on angular momentum ($J$) (neutron ($N$) and proton ($Z$) numbers) is provided in the SM. The complex number $c_0$ ensures the normalization of the projected state.

To characterize and quantify the difference between the QR and CRR analysis of ultra-relativistic ion-ion collisions, controlled approximations to the PHFB normalized quadrupole mean-square eccentricity, and associated effective deformation parameter $\mathcal{B}^2_{2}(\text{HE})_{\text{QR}}$, must be performed. As detailed in the SM, the CRR description characterized by Eqs.~\eqref{kbodydensityRR}-\eqref{eccCRR} can indeed be formally recovered from PHFB theory via two successive approximations
\begin{enumerate}
\item the so-called {\it needle} approximation to the symmetry projection neglects the quantum entanglement in the angular-momentum projected many-body wave-function to deliver the quantum {\it rigid} rotor (QRR) approximation $\mathcal{B}^2_{2}(\text{HE})_{\text{QRR}}$ as an intermediate step. From a formal perspective, this approximation amounts to considering that the overlap between rotated mean-field states reduces to a delta distribution, i.e., $\langle \Phi(\Omega') | \Phi(\Omega) \rangle \approx \delta(\Omega'-\Omega)$. Therefore,
this approximation relates to a geometrical argument whose effect is likely to depend on the nuclear mass and intrinsic deformation.
\item further ignoring the quantum (i.e., exchange and pairing) contributions to $\mathcal{B}^2_{2}(\text{HE})_{\text{QRR}}$ associated with the fermionic character of nucleons eventually delivers the CRR value.
\end{enumerate}

\paragraph{Numerical results.}
\nocite{frosini26a}
Projected HFB calculations are presently performed within two different frameworks. First, PHFB calculations are performed within the {\it ab initio} theoretical scheme~\cite{Yao:2019rck,Frosini2022a,Frosini2022b} based on the EM1.8/2.0~\cite{Hebeler11a} Chiral effective field theory ($\chi$EFT) Hamiltonian for five even-even nuclei ($^{20}\mathrm{Ne}$, $^{28}\mathrm{Si}$,  $^{40}\mathrm{Ca}$, $^{76}\mathrm{Ge}$ and $^{136}\mathrm{Xe}$). These representative isotopes span different mass regions as well as deformation regimes and contain possible candidates for future experimental ion-ion runs \cite{lightions2025}. A constraint is placed on $Q_{20}[\text{dHFB}]$ to span a large range of intrinsic deformations in each nucleus~\footnote{The dHFB and PHFB energies computed in $^{76}\mathrm{Ge}$ and $^{136}\mathrm{Xe}$ beyond $\beta_{20}[\text{dHFB}]=0.4$ is compromised due to a collapse of the rank-reduction method of the three-nucleon interaction~\protect\cite{frosini26a}. Still, this issue has no consequence on the computed intrinsic and effective quadrupole deformations of present interest.}. In this context, PHFB provides the QR approximation to the exact solutions of Schrodinger's equation. Second, systematic results over a very large set of nuclei up to transactinides ($Z\in[4,100]$) can be accessed via PHFB calculations performed within the multi-reference energy density functional (MR-EDF) framework~\cite{Yao:2010,Yao:2014} based on the PC-F1  covariant EDF parameterization~\cite{Burvenich:2002}. In this case, the intrinsic deformation $\beta^2_{20}[\text{dHFB}]$ corresponds to the minimum of the intrinsic dHFB energy. Details about the numerical parameters used to perform both sets of calculations are provided in the SM. Additional {\it ab initio} and MR-EDF results are  discussed in the SM to complement the analysis proposed in this Letter.

\begin{figure}[htbp]     
  \centering             
  \includegraphics[width=0.4\paperwidth]{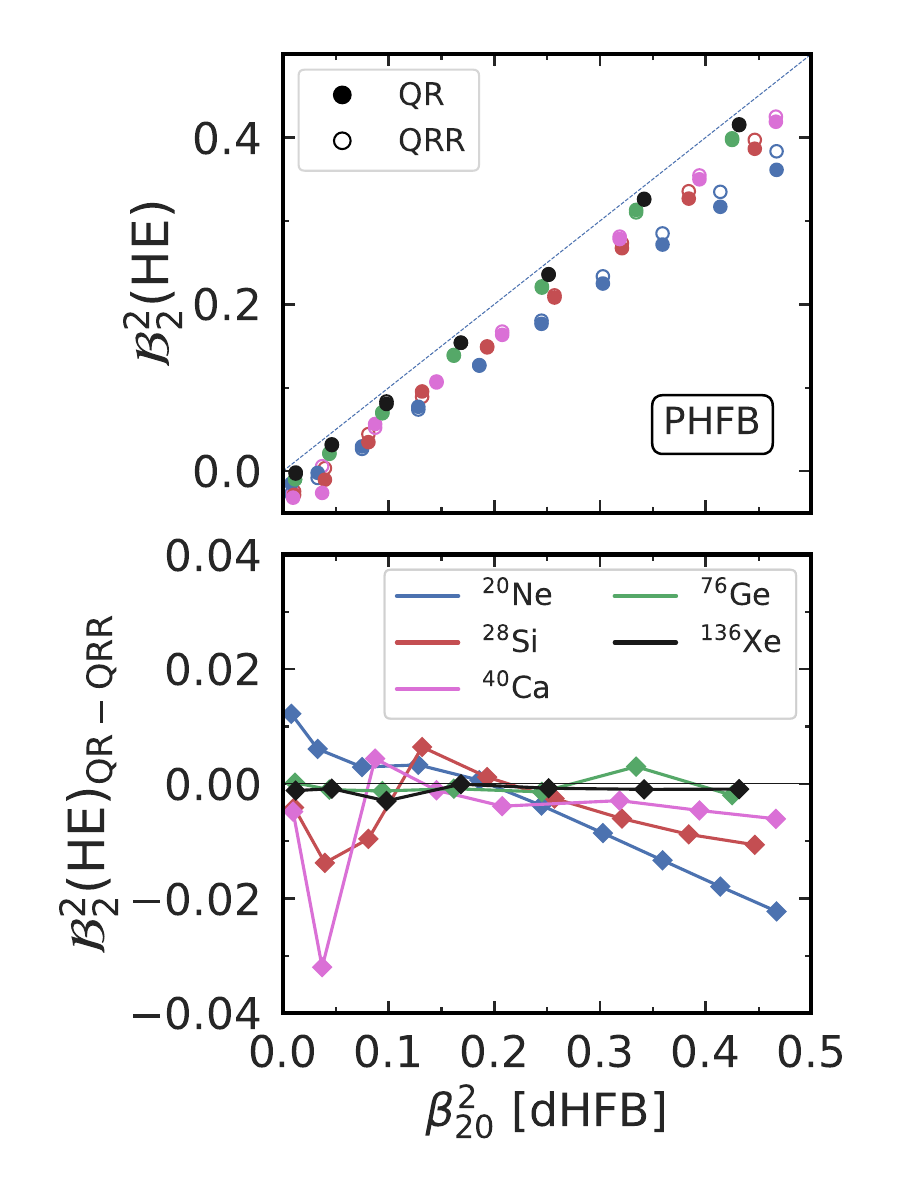}
  \caption{Effective deformation ${\cal B}_2^2(\mathrm{HE})$ against $\beta^2_{20}[\text{dHFB}]$ for $^{20}\mathrm{Ne}$, $^{28}\mathrm{Si}$,  $^{40}\mathrm{Ca}$, $^{76}\mathrm{Ge}$ and $^{136}\mathrm{Xe}$ computed within the QR (i.e. with exact projection) and the QRR (i.e. with the needle approximations) approximations using the EM1.8/2.0 $\chi$EFT Hamiltonian. Upper panel: absolute values (the dashed line shows the first diagonal). Lower panel: absolute error of the QR with respect to the QRR.}
  \label{fig:B2sq_rigid_quant_beta2sq}
\end{figure}

The top panel of Fig.~\ref{fig:B2sq_rigid_quant_beta2sq} displays the QR and QRR effective quadrupole deformations ${\cal B}_2^2(\mathrm{HE})$ of  $^{20}\mathrm{Ne}$,  $^{28}\mathrm{Si}$,  $^{40}\mathrm{Ca}$, $^{76}\mathrm{Ge}$ and $^{136}\mathrm{Xe}$ as a function of the constrained $\beta^2_{20}[\text{dHFB}]$~\footnote{Even though the absolute minimum in $^{28}\mathrm{Si}$ is oblate, only prolate configurations with $\beta^2_{20}[\text{dHFB}]\geq 0$ are plotted in Fig.~\ref{fig:B2sq_rigid_quant_beta2sq}. Had oblate configurations been used, the results and conclusions would have remained the same.}.  As identified in Ref.~\cite{Bofos:2026huw}, ${\cal B}_2^2(\mathrm{HE})_{\text{QR}}$ is strongly correlated with $\beta^2_{20}[\text{dHFB}]$ but  is not equal to it. ${\cal B}_2^2(\mathrm{HE})_{\text{QR}}$ becomes even negative for $\beta^2_{20}[\text{dHFB}] \lesssim 0.03$ in light nuclei. The offset with respect to the first diagonal is seen to decrease with $A$. 

The comparison between ${\cal B}_2^2(\mathrm{HE})_{\text{QR}}$ and its QRR counterpart proves that the error induced by the needle approximation to the angular momentum projection is  small overall~\footnote{When constraining over such a large span of $\beta^2_{20}[\text{dHFB}]$ values, the needle approximation to ${\cal B}_2^2(\mathrm{HE})_{\text{QR}}$ is seen to become abruptly unsafe for nuclei with $A \leq 16$.}. As the bottom panel of Fig.~\ref{fig:B2sq_rigid_quant_beta2sq} illustrates, this error evolves non-trivially with both $A$ and $\beta^2_{20}[\text{dHFB}]$~\footnote{The impact of the needle approximation, and so its behavior as a function of both $\beta^2_{20}[\text{dHFB}]$ and $A$, cannot be anticipated and depends on the system and observable under consideration. This is illustrated in the SM by comparing the impact of the needle approximation on ${\cal B}_2^2(\mathrm{HE})$, the total energy $E$ and the total angular momentum $J^2$.}. The maximum error is seen under the form of a spike around $\beta^2_{20}[\text{dHFB}]=0.04$ ($\beta_{20}[\text{dHFB}]=0.2$) in $^{40}\mathrm{Ca}$~\footnote{This trend of the error with respect to $\beta^2_{20}[\text{dHFB}]$ is common to all doubly-magic nuclei with $A > 16$. The amplitude of the spike is seen to decrease with both $A$ and isospin asymmetry.}.  

Eventually, systematic PHFB calculations performed at the intrinsic deformation of the HFB minimum show that the impact of the needle approximation on the effective quadrupole deformation remains smaller than $10\%$ for the large majority of nuclei, except for a few light systems where it reaches $30\%$; see the SM for details. To this level of accuracy, the above analysis validates the ``traditional'' interpretation of ultra-relativistic ion-ion collisions. However, this justification does not rely on the (incorrect) time-scale argument usually put forward~\cite{Dobaczewski:2025rdi}. This conclusion is consistent with the analysis conducted in Ref.~\cite{Ke:2025}.

\begin{figure}[htbp]     
  \centering             
  \includegraphics[width=0.4\paperwidth]{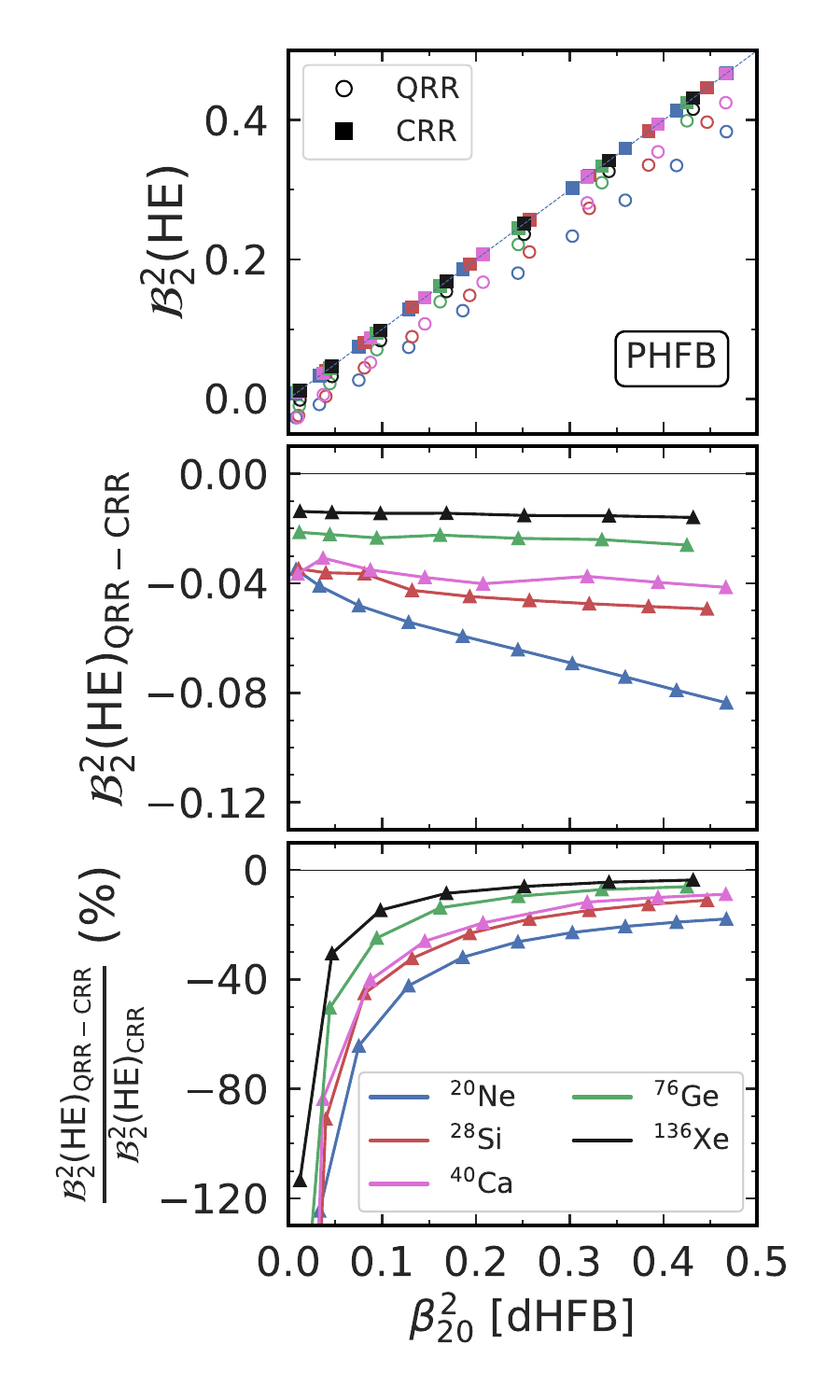}
  \caption{Effective deformation ${\cal B}_2^2(\mathrm{HE})$ against $\beta^2_{20}[\text{dHFB}]$ of $^{20}\mathrm{Ne}$, $^{28}\mathrm{Si}$,  $^{40}\mathrm{Ca}$, $^{76}\mathrm{Ge}$ and $^{136}\mathrm{Xe}$ computed based on the needle approximation using the EM1.8/2.0 $\chi$EFT Hamiltonian. Upper panel: QRR (i.e. including direct, exchange and pairing terms) and CRR (i.e. including only the direct term) values. The dashed line shows the first diagonal. Middle panel: contribution from the exchange and pairing terms to the QRR effective deformation. Lower panel: relative exchange and pairing contributions ($\%$) to the CRR effective deformation.}
  \label{fig:dhfb_B2_sq_beta2_sq}
\end{figure}

The upper panel of Fig.~\ref{fig:dhfb_B2_sq_beta2_sq} compares ${\cal B}_2^2(\mathrm{HE})$ computed within the QRR and CRR approximations as a function of $\beta^2_{20}[\text{dHFB}]$ for the five selected nuclei. Their difference originates from the quantum contribution to the two-body part of the QRR mean-square eccentricity associated with the so-called exchange (Fock) and pairing (Bogoliubov) terms. The present results prove that removing these contributions brings the effective deformation back on the diagonal, i.e., it shows that $\mathcal{B}_2^2(\mathrm{HE})_{\text{CRR}}=\beta^2_{20}[\text{dHFB}]$ to very good approximation. The systematic analysis provided in the SM demonstrates that such an equality actually holds within $5\%$ as soon as $\beta^2_{20}[\text{dHFB}] \geq 0.03$ ($\beta_{20}[\text{dHFB}] \geq 0.17$), the agreement further improving monotonically with $\beta^2_{20}$. The middle panel of Fig.~\ref{fig:dhfb_B2_sq_beta2_sq} singles out the subtracted contributions, showing that their sum is negative, grows linearly but mildly with $\beta^2_{20}[\text{dHFB}]$ and decreases monotonically with $A$. The lower panel  of Fig.~\ref{fig:dhfb_B2_sq_beta2_sq} shows that the quantum contributions dominate the QRR effective quadrupole deformation in light nuclei and/or nuclei displaying a small or moderate intrinsic deformation. These  contributions drop below $10\%$ of the classical contribution $\beta^2_{20}[\text{dHFB}]$ in relatively heavy ($A\geq 100$) well-deformed nuclei ($\beta^2_{20}[\text{dHFB}] \geq 0.15$, i.e. $\beta_{20}[\text{dHFB}] \geq 0.4$).

\begin{figure}[tb]     
  \centering             
  \includegraphics[width=\columnwidth]{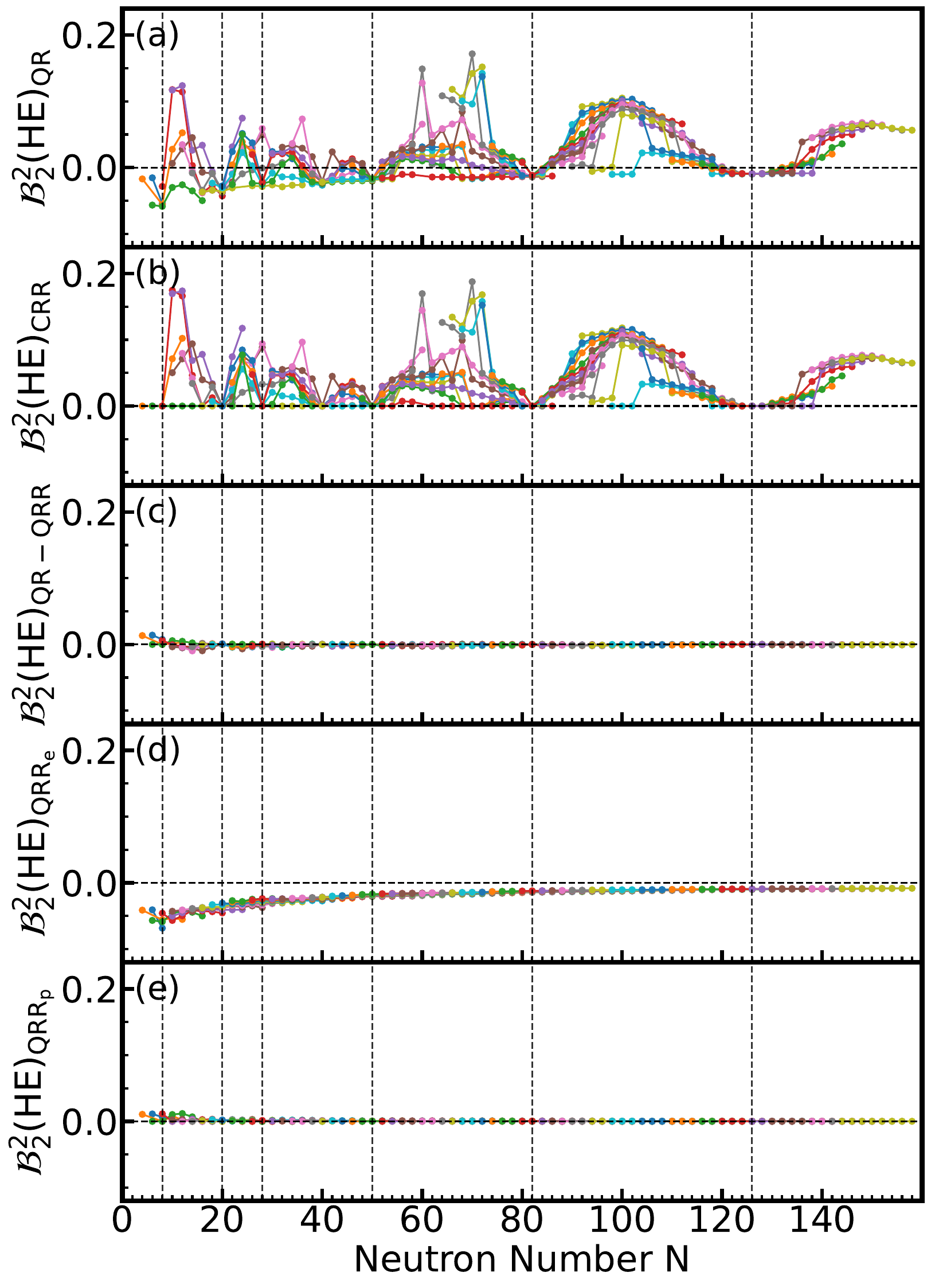}
  \caption{Systematic of effective quadrupole deformation as a function of neutron number from PHFB calculations of isotonic chains with $Z \in [4,100]$ performed within the MR-EDF framework. Vertical dashed lines indicate the traditional magic numbers inferred from stable nuclei. (a) quantum rotor, (b) classical rigid rotor, (c) difference between the exact symmetry restoration and its needle approximation, (d) exchange contribution in the needle approximation (e) pairing contribution in the needle approximation.}
  \label{Figsystematic}
\end{figure}

Figure~\ref{Figsystematic} extends the previous discussion to a systematic over numerous isotonic chains ($Z \in [4,100]$), all the way to transactinides, based on PHFB calculations performed within the MR-EDF theoretical scheme \cite{Yao:2010,Yao:2014}. Panel (a)  illustrates that ${\cal B}_2^2(\mathrm{HE})_{\text{QR}}$ displays strong shell effects with marked minima at expected neutron magic numbers and maxima in between~\footnote{Two isotopic chains ($Z=14$ and $16$) do not display minima at $N=28$, indicating the disappearance of such a neutron magic number in proton deficient isotones.}. However, the baseline value at those neutron magic numbers and along proton semi-magic chains is seen to be negative. Panel (b) demonstrates that  ${\cal B}_2^2(\mathrm{HE})_{\text{CRR}}$ retains the main patterns displayed in the QR while removing the negative baseline. In doing so, the systematic indeed matches the behavior expected from the  intrinsic axial quadrupole deformation $\beta^2_{20}[\text{dHFB}]$ in nuclei. Panel (c) confirms that, while the full quantum angular-momentum projection is necessary for a quantitative extraction of the effective deformation in light nuclei, the needle approximation becomes quickly justified as $A$ increases. Panel (d) clarifies that the negative offset in the upper panel is due to the exchange contribution associated with Pauli's exclusion principle. Interestingly, this contribution is insensitive to shells effects and thus behaves qualitatively differently from $\beta^2_{20}[\text{dHFB}]$ with respect to $N$. Starting from a maximum (absolute) value around $0.06$ in light nuclei, it decreases monotonically with $A$~\footnote{The exchange term converges to zero in the $A\rightarrow \infty$ limit, thus showing that the QR becomes fully classical in this limit.}. This exchange contribution accounts for $100\%$ of the effective deformation in nuclei with $\beta_{20}[\text{dHFB}]\approx0$. As shown in the SM, it drops below $10\%$ in well deformed nuclei, i.e., when $\mathcal{B}^{2}_{2}(\text{HE})_{\text{QR}}$ becomes larger than about $0.14$.  Panel (e) eventually concludes that the pairing contribution to the QRR effective deformation is negligible. 

\paragraph{Summary.} 
In this work, quantum effects arising in a quantum rotor analysis of symmetric ultra-central ultra-relativistic ion-ion collisions are quantified for the first time. Doing so, the approximate treatment of the angular momentum restoration at play in the traditional interpretation based on the classical rigid rotor picture is validated within a $10\%$ error on the extracted effective deformation. Interestingly, the reason why this approximation works well is at variance with the improper time-scale argument that has been put forward so far~\cite{Dobaczewski:2025rdi}. The classical rigid rotor is also shown to omit a second type of quantum contributions associated with the fermionic character of the nucleons. These contributions are shown to be non negligible in light nuclei and to be independent of shell effects, as well as to decrease quickly and monotonically with $A$. Eventually, the present analysis constitutes an important step towards a quantum-mechanically meaningful extraction of the intrinsic nuclear deformation through symmetric ultra-central ultra-relativistic ion-ion collisions. The next step will consist of including dynamical correlations that are yet missing in the quantum rotor description of the collided nuclei in order to perform a fully quantitative extraction of the intrinsic deformation.

\bigskip
\paragraph{Acknowledgements.} 
We thank the organizers and all the participants of the YITP workshop \textit{``Intersection of nuclear structure and high-energy nuclear collisions 2026''} for useful discussions. The calculations were partly performed using computational resources from CCRT (TOPAZE supercomputer). This work has received funding from the European Research Council under the European Union’s Horizon Europe Research and Innovation Programme (Grant Agreement No.\ 101162059). Y. L., C.R. D. and J.M. Y. are supported in part by the National Natural Science Foundation of China (Grant Nos.  125B2108, 12405143, and  12375119).


\section{Supplemental Material}


\section{Mean-square azimuthal flow vs eccentricity}

The anisotropy of the azimuthal hadronic flow in the transverse plane of ultra-relativistic ion-ion collisions is characterized by Fourier coefficients $V_{\ell}$ of multipolarity $\ell$. This anisotropic flow is further related, through hydrodynamic evolution of the QGP, to the normalized eccentricities $\epsilon_{\ell}$  of the entropy density $s(\mathbf{r}_{\perp})$ deposited shortly after the reaction~\cite{Sousa:2024msh}
\begin{equation}
   \epsilon_{\ell} \equiv -\frac{\mathcal{E}^{(1)}_{\ell}}{R^{(1)}_{\ell}} \equiv - \frac{\int_{\mathbf{r}_{\perp}} \mathcal{E}^{(1)}_{\ell}(\mathbf{r}_{\perp}) s(\mathbf{r}_{\perp})}{\int_{\mathbf{r}_{\perp}} R^{(1)}_{\ell}(\mathbf{r}_{\perp}) s(\mathbf{r}_{\perp})} \, , \label{SM_eccentricities}
\end{equation}
where the coordinate vector is given by
\begin{equation}
\mathbf{r} = (\mathbf{r}_{\perp},z) = (r_{\perp},\phi,z)  \, ,
\end{equation}
such that $\mathbf{r}_{\perp}$ denotes the position vector in the transverse plane whereas $\phi$ specifies the corresponding azimuthal angle. In Eq.~(\ref{SM_eccentricities}), the eccentricity and transverse radius are obtained from
\begin{subequations}
\begin{align}
 \mathcal{E}^{(1)}_{\ell}(\mathbf{r}_{\perp}) &\equiv r^\ell_{\perp} e^{i\ell \phi} = r^\ell \sin^\ell \theta e^{i\ell \phi} \, , \\
 R^{(1)}_{\ell}(\mathbf{r}_{\perp}) &\equiv r^\ell_{\perp} = r^\ell \sin^\ell \theta  \, ,
\end{align}
 \end{subequations}
where $(r,\theta,\phi)$ denote spherical coordinates. 

The anisotropy of the azimuthal flow between two particles in the final state is given by the variance of $V_{\ell}$. In ultra-central collisions, such a variance is, to a good approximation, shown to be proportional to the normalized variance of the sole eccentricity of identical multipolarity~\cite{Niemi:2016}, i.e.,
\begin{equation}
\text{Var}(V_{\ell}) \approx \kappa^2_{\ell}  \text{Var}(\epsilon_{\ell}) \, .
\end{equation}

In symmetric collisions, $\text{Var}(\epsilon_{\ell})$  was further shown to be related to the quantum mechanical expectation value of the squared eccentricity operator $\mathcal{E}^{(2)}_{\ell}$ in the $J=0$ ground state $| \Psi^{0^+}\rangle$ of the colliding nuclei according to~\cite{Duguet2025a}
\begin{align}
    \text{Var}(\epsilon_{\ell})
    &\equiv \langle \delta \epsilon^{2}_{\ell}   \rangle \nonumber \\
    &= \frac{1}{2} \frac{\langle \Psi^{0^+}  | \mathcal{E}^{(2)}_{\ell}| \Psi^{0^+}\rangle }{\langle   \Psi^{0^+}  |  R^{(1)}_{\ell} | \Psi^{0^+}  \rangle^2} \label{SM_meansquareeccentricity}  \\
    &= \frac{1}{2} 
    \frac{\int_{\mathbf{r}_{1}} \mathcal{E}^{(2) \, (1\text{b})}_{\ell}(\mathbf{r}_{1}) \rho^{(1)}(\mathbf{r}_{1}) +\int_{\mathbf{r}_{1,2}}  \mathcal{E}^{(2) \,  \, (2\text{b})}_{\ell}(\mathbf{r}_{1},\mathbf{r}_{2}) \rho^{(2)}(\mathbf{r}_{1}, \mathbf{r}_{2})}{\Big[ \int_{\mathbf{r}_{1}} R^{(1)}_{\ell}(\mathbf{r}_{1}) \rho^{(1)}(\mathbf{r}_{1}) \Big]^2} \, ,\nonumber
\end{align}
where the operator $\mathcal{E}^{(2)}_{\ell}$ has been separated into its one-body and two-body components given by
\begin{subequations}
\begin{align}
\mathcal{E}^{(2) \, (1\text{b})}_{\ell}(\mathbf{r}_{1}) &\equiv  \mathcal{E}^{(1)}_{\ell}(\mathbf{r}_{1}) \mathcal{E}^{(1)}_{-\ell}(\mathbf{r}_{1}) \, ,\\
\mathcal{E}^{(2) \,  \, (2\text{b})}_{\ell}(\mathbf{r}_{1},\mathbf{r}_{2}) & \equiv \mathcal{E}^{(1)}_{\ell}(\mathbf{r}_{1}) \mathcal{E}^{(1)}_{-\ell}(\mathbf{r}_{2}) \, .
\end{align}
\end{subequations}
and where the ground-state local one and two-body densities are defined as\footnote{The even-even nuclear ground state $| \Psi^{0^+} \rangle$ is taken to be normalized.}
\begin{subequations}
\label{1and2bodydensities}
\begin{align}
\rho^{(1)}(\mathbf{r}_{1}) &\equiv \langle \Psi^{0^+} | a^{\dagger}(\mathbf{r}_{1}) a(\mathbf{r}_{1}) | \Psi^{0^+} \rangle \, , \\
\rho^{(2)}(\mathbf{r}_{1},\mathbf{r}_{2})  &\equiv \langle \Psi^{0^+} | a^{\dagger}(\mathbf{r}_{1}) a^{\dagger}(\mathbf{r}_{2}) a(\mathbf{r}_{2})a(\mathbf{r}_{1}) | \Psi^{0^+} \rangle \, .
\end{align}
\end{subequations}

\section{Classical rigid rotor}

\subsection{Rigid rotor}

One way to formulate the {\it rigid} rotor (RR) model is to stipulate that the laboratory $k$-body density $\rho^{(k)}_{\text{RR}}$ of a given state is obtained by averaging over all spatial orientations an intrinsic $k$-body density $\rho^{(k)}_{\Omega}$ pointing in direction $\Omega\equiv (\alpha,\beta,\gamma)$ with appropriate weights given by Wigner functions; e.g.,the RR one- and two-body local densities of a $J=0$ state read as
\begin{subequations}
\label{SM_kbodydensityRR}
\begin{align}
\rho^{(1)}_{\text{RR}}(\mathbf{r}_1)
         &= \frac{1}{8\pi^2} \int_{\Omega}  
       \rho^{(1)}_{\Omega}\big(\mathbf{r}_1\big) \, , \label{SM_kbodydensityRR1}
       \\
\rho^{(2)}_{\text{RR}} (\mathbf{r}_1,\mathbf{r}_2)
        &= \frac{1}{8\pi^2} \int_{\Omega}  
       \rho^{(2)}_{\Omega} (\mathbf{r}_1,\mathbf{r}_2)  \, , \label{SM_kbodydensityRR2}
\end{align}
\end{subequations}
such that the former only depends on $r_1$, and is thus spherically symmetric, whereas the latter  depends on $r_1,r_2$ and the relative angle $\cos\zeta \equiv \mathbf{r}_1 \cdot \mathbf{r}_2/r_1 r_2$.

\subsection{Classical rigid rotor}

The {\it classical} RR (CRR) further postulates that the system displays no correlations {\it in the intrinsic frame} such that the intrinsic $k$-body density is simply the product of $k$ 1-body densities, e.g., the local two-body density then reads
\begin{align}
\rho^{(2)}_{\Omega} (\mathbf{r}_1,\mathbf{r}_2)
        &\equiv
       \rho^{(1)}_{\Omega}\big(\mathbf{r}_1\big) \, 
       \rho^{(1)}_{\Omega}\big(\mathbf{r}_2\big)  \, . \label{SM_nocorrintrinsic}
\end{align}
In this CRR, correlations in the laboratory $k$-body density are entirely induced by the rotation of a deformed intrinsic $k$-body density.

\subsection{Normalized quadrupole mean-square eccentricity}
\label{secCRR}
\nocite{Mehrabpour:2026yuc}
 
Among simple models of the one-body intrinsic density $\rho^{(1)}_{\Omega=0}(\mathbf{r})$, one convenient parametrization is given by~\cite{Duguet2025a}
\begin{align}
    \rho^{(1)}_{\Omega=0}(\mathbf{r}) &\equiv \frac{Ae^{-r^2/2{\cal R}(\theta,\phi)^2}}{(2\pi)^{3/2}R^{3}} \, , \label{param1bodyintrinsicdens1}
\end{align}
where $A$ denotes the total number of nucleons and where the axially deformed surface radius is expressed in terms of the axial quadrupole intrinsic deformation parameter  $\beta_{20}$ through\footnote{The present discussion can be easily extended to octupole deformation and/or triaxial shapes; see Refs.~\protect\cite{Duguet2025a} and~\protect\cite{Mehrabpour:2026yuc}, respectively.}
\begin{equation}
    {\cal R}(\theta,\phi) \equiv R[1 + \beta_{20} Y_2^0(\theta,\phi) 
    ] \, , \label{param1bodyintrinsicdens2}
\end{equation}
with $R$ being a constant.

For such a particular model, analytical expressions of $\rho^{(1)}_{\text{CRR}}(\mathbf{r}_1)$ and $\rho^{(2)}_{\text{CRR}} (\mathbf{r}_1,\mathbf{r}_2)$ can be obtained; see Ref.~\cite{Duguet2025a} for details. Inserting these expressions into Eq.~(\ref{SM_meansquareeccentricity}) delivers the quadrupole ($\ell=2$) mean-square eccentricity of the axial CRR as~\cite{Duguet2025a}
\begin{subequations}
\begin{align}
    \langle \delta \epsilon^2_2 \rangle^{\text{CRR}}_{0^+} &= \frac{1}{A} + \frac{3}{4\pi} \beta^2_{20} \, .
\end{align}
\end{subequations}
The one-body contribution is proportional to $A^{-1}$ and independent of the intrinsic deformation parameter. The two-body contribution probing genuine two-body correlations is independent of $A$ but scales with the square of the intrinsic deformation parameter.

\subsection{Effective axial quadrupole deformation}

Based on this CRR model result, a quantity homogeneous to the square of a dimensionless axial quadrupole deformation parameter is introduced for an arbitrary $J^\pi=0^+$ state $| \Psi^{0^+} \rangle$ according to~\cite{Blaizot2025a}
\begin{equation}
    \label{eq: SM Beta parameter definition1}
    \mathcal{B}^2_{2}(\text{HE}) \equiv \frac{4\pi}{3} \langle  \delta \epsilon^{2 \, (2\text{b})}_2 \rangle \; .
\end{equation}
By definition, $\mathcal{B}^2_{2}(\text{HE})$ reduces to the square of the intrinsic axial quadrupole deformation in the CRR model, i.e., $\mathcal{B}^{2}_{2}(\text{HE})_{\text{CRR}} = \beta^2_{20}$.

\section{Quantum rotor}

\subsection{Projected HFB approximation}

In actual experiments, the collided nuclei are of course quantum systems. Consequently, the present study focuses on results obtained from the quantum rotor (QR) approximation provided by projected Hartree-Fock Bogoliubov (PHFB) calculations. This approximation relies on the following ansatz for the $J^\pi=0^+$ ground-state of even-even nuclei
\begin{equation}
| \Theta^{0^+}_{\text{PHFB}} \rangle \equiv \frac{1}{c_{0}} P^{J=0} P^{\text{N}\text{Z}}  | \Phi \rangle \, , \label{SM_PHFBansatz}
\end{equation}
where the normalized symmetry-breaking Hartree-Fock Bogoliubov (HFB) state $| \Phi \rangle$ displays the axial quadrupole deformation given by
\begin{align}
\beta_{20} &\equiv \frac{4\pi}{5} \frac{\langle \Phi | Q_{20} | \Phi \rangle}{\langle  \Phi | r^2 |  \Phi \rangle}
\, ,
\end{align}
with
\begin{equation}
    \label{eq: SM quadupole operator definition}
    Q_{\ell \mu} \equiv \sum_{i=1}^{A} r^\ell_{i} Y^{\mu}_{\ell}(\Omega_{i})\; .
\end{equation}
In Eq.~\eqref{SM_PHFBansatz}, the normalization constant is given by
\begin{equation}
    \label{normalization}
c^2_{0} = \langle \Phi |P^{0} P^{\text{N}\text{Z}} | \Phi \rangle  \, ,
\end{equation}
where the hermitian and idempotent character of the projection operators was employed. 

The restorations of angular momentum ($J=0$), as well as of neutron ($N$) and proton ($Z$) numbers, proceed via the application of the operators
\begin{subequations}
\label{projectors}
\begin{align}
P^{\text{J}} &\equiv \frac{2J+1}{8\pi^2} \int_{[0,2\pi]\times[0,\pi]\times[0,2\pi]} \hspace{-1.5cm} D^{\text{J}*}_{00}(\Omega) \, R_{J}(\Omega) \, , \\
P^{\text{N}} &\equiv \frac{1}{2\pi}\int_{[0,2\pi]}  d\varphi_N \, e^{i\varphi_n \text{N}} R_{N}(\varphi_n) \, , \\
P^{\text{Z}} &\equiv \frac{1}{2\pi}\int_{[0,2\pi]}  d\varphi_Z \, e^{i\varphi_z \text{Z}} R_{Z}(\varphi_z) \, , \\
P^{\text{NZ}} &\equiv  P^{\text{N}} P^{\text{Z}} \, , 
\end{align}
\end{subequations}
where $D^{\text{J}}_{MK}(\Omega)$ denotes Wigner functions. The rotation operators are defined as
\begin{subequations}
\label{rotation_operators}
\begin{align}
R_{J}(\Omega) &\equiv e^{-i\alpha J_z} e^{-i\beta J_y} e^{-i\gamma J_z} \, , \\
R_{N}(\varphi_n) &\equiv e^{-i\varphi_n N} \, , \\
R_{Z}(\varphi_z) &\equiv e^{-i\varphi_z Z} \, , 
\end{align}
\end{subequations}
with $J_{i=y,z}$, $N$ and $Z$ are the usual angular momentum $i$-component, neutron and proton number one-body operators, respectively. To make analytical expressions more compact, the projections on $N$ and $Z$ are replaced by a single projection on the particle number $A$ parametrized by the angle $\varphi$ in the following. 


\subsection{One- and two-body local densities}

Based on Eqs.~\eqref{1and2bodydensities} and~\eqref{SM_PHFBansatz}, QR one- and two-body local densities are given by
\begin{subequations}
\label{PHFBdens}
\begin{align}
\rho^{(1)}_{\text{QR}}(\mathbf{r}_{1}) &\equiv \langle \Theta^{0^+}_{\text{PHFB}} | a^{\dagger}(\mathbf{r}_{1}) a(\mathbf{r}_{1}) | \Theta^{0^+}_{\text{PHFB}} \rangle \nonumber\\
&=  \frac{1}{(8\pi^2c_{0})^2} \int_{\Omega'\Omega}\rho^{(1)}_{\Omega'\Omega}(\mathbf{r}_{1}) \, , \label{PHFB1bodydensbody} \\
\rho^{(2)}_{\text{QR}}(\mathbf{r}_{1},\mathbf{r}_{2}) &\equiv \langle \Theta^{0^+}_{\text{PHFB}} | a^{\dagger}(\mathbf{r}_{1}) a^{\dagger}(\mathbf{r}_{2}) a(\mathbf{r}_{2}) a(\mathbf{r}_{1}) | \Theta^{0^+}_{\text{PHFB}} \rangle \nonumber\\
&=  \frac{1}{(8\pi^2c_{0})^2} \int_{\Omega'\Omega} \rho^{(2)}_{\Omega'\Omega}(\mathbf{r}_{1},\mathbf{r}_{2})  \, , \label{PHFB2bodydensbody}
\end{align}
\end{subequations}
where their off-diagonal intrinsic counterparts are respectively defined as
\begin{subequations}
\label{offdiagPHFBdensity}
\begin{align}
\rho^{(1)}_{\Omega'\Omega}(\mathbf{r}_{1}) \equiv& \langle \Phi (\Omega') | a^{\dagger}(\mathbf{r}_{1})a(\mathbf{r}_{1})  P^{A}| \Phi (\Omega) \rangle \nonumber \\
=& \frac{1}{2\pi}  \int_{[0,2\pi]}  \!\! d\varphi \, e^{i\varphi \text{A}} \, \langle \Phi(\Omega') | \Phi(\Omega,\varphi) \rangle \nonumber \\
& \hspace{1.2cm} \times \rho^{(1)}_{\Omega'\Omega\varphi}(\mathbf{r}_{1},\mathbf{r}_{1}) \, , \label{1bodyoffdiagPHFBdensity}
\end{align}
and as
\begin{align}
\rho^{(2)}_{\Omega'\Omega}(\mathbf{r}_{1},\mathbf{r}_{2}) \equiv& \langle \Phi (\Omega') | a^{\dagger}(\mathbf{r}_{1}) a^{\dagger}(\mathbf{r}_{2}) a(\mathbf{r}_{2}) a(\mathbf{r}_{1})  P^{A}| \Phi (\Omega) \rangle \nonumber \\
=& \frac{1}{2\pi}  \int_{[0,2\pi]} \!\! d\varphi \, e^{i\varphi \text{A}} \, \langle \Phi(\Omega') | \Phi(\Omega,\varphi) \rangle \nonumber \\
& \hspace{0.9cm} \times \left[\rho^{(1)}_{\Omega'\Omega\varphi}(\mathbf{r}_{1},\mathbf{r}_{1})\rho^{(1)}_{\Omega'\Omega\varphi}(\mathbf{r}_{2},\mathbf{r}_{2}) \right. \nonumber \\
& \hspace{1.2cm}  -\rho^{(1)}_{\Omega'\Omega\varphi}(\mathbf{r}_{2},\mathbf{r}_{1})\rho^{(1)}_{\Omega'\Omega\varphi}(\mathbf{r}_{1},\mathbf{r}_{2}) \nonumber \\
&\hspace{1.2cm} \left.+ \bar{\kappa}^{(1)\ast}_{\Omega'\Omega\varphi}(\mathbf{r}_{1},\mathbf{r}_{2})\kappa^{(1)}_{\Omega'\Omega\varphi}(\mathbf{r}_{2},\mathbf{r}_{1})\right]  \, . \label{2bodyoffdiagPHFBdensity}
\end{align}
\end{subequations}
In Eqs.~\eqref{1bodyoffdiagPHFBdensity}-\eqref{2bodyoffdiagPHFBdensity}, the doubly rotated Bogoliubov state is defined as
\begin{equation}
| \Phi(\Omega,\varphi) \rangle \equiv R_{J}(\Omega) R_{A}(\varphi)| \Phi \rangle \, ,
\end{equation}
such that $| \Phi(\Omega) \rangle \equiv | \Phi(\Omega,0) \rangle$. Furthermore, the fully off-diagonal intrinsic one-body local densities are given by
\begin{subequations}
\label{offdiagonal1bodydens}
\begin{align}
\rho^{(1)}_{\Omega'\Omega\varphi}(\mathbf{r}_{1},\mathbf{r}_{2}) &\equiv \frac{\langle \Phi(\Omega') | a^{\dagger}(\mathbf{r}_{2})a(\mathbf{r}_{1}) | \Phi(\Omega,\varphi) \rangle }{\langle \Phi(\Omega') | \Phi(\Omega,\varphi) \rangle } \, , \\
\kappa^{(1)}_{\Omega'\Omega\varphi}(\mathbf{r}_{1},\mathbf{r}_{2}) &\equiv \frac{\langle \Phi(\Omega') | a(\mathbf{r}_{2})a(\mathbf{r}_{1}) | \Phi(\Omega,\varphi) \rangle }{\langle \Phi(\Omega') | \Phi(\Omega,\varphi) \rangle } \, , \\
\bar{\kappa}^{(1)\ast}_{\Omega'\Omega\varphi}(\mathbf{r}_{1},\mathbf{r}_{2}) &\equiv \frac{\langle \Phi(\Omega') | a^{\dagger}(\mathbf{r}_{1})a^{\dagger}(\mathbf{r}_{2}) | \Phi(\Omega,\varphi) \rangle }{\langle \Phi(\Omega') | \Phi(\Omega,\varphi) \rangle } \, . 
\end{align}
\end{subequations}
The three terms in Eq.~\eqref{2bodyoffdiagPHFBdensity}, denoted as the direct (Hartree), exchange (Fock) and pairing (Bogoliubov) contributions, result from the application of the off-diagonal Wick's theorem~\cite{Balian:1969}.

\subsection{Two-body mean-square eccentricity}

The QR local densities of Eq.~\eqref{PHFBdens} are eventually inserted into Eq.~\eqref{SM_meansquareeccentricity} to compute QR one- and two-body contributions to the normalized mean-square eccentricity of multipolarity $\ell$. From the two-body contribution, the QR effective quadrupole deformation parameter $\mathcal{B}^2_{2}(\text{HE})_{\text{QR}}$ is obtained through Eq.~\eqref{eq: SM Beta parameter definition1}.

Focusing on this two-body contribution, one can write it as 
\begin{align}
\langle \Theta^{0^+}_{\text{PHFB}}  | \mathcal{E}^{(2)(2b)}_{\ell}| \Theta^{0^+}_{\text{PHFB}}\rangle 
    &= \int_{\mathbf{r}_{1,2}} \! \mathcal{E}^{(2) \,  \, (2\text{b})}_{\ell}(\mathbf{r}_{1},\mathbf{r}_{2}) \, \rho^{(2)}_{\text{QR}}(\mathbf{r}_{1}, \mathbf{r}_{2}) \, . \label{eccPHFB}
\end{align}
Because the expectation value involves a $J^\pi=0^+$ state, only the scalar component ($L=0$) obtained through the multipole expansion of the operator~\cite{Bofos:2026huw}
\begin{equation}
\label{eq: SM 2-body eccentricity operator decomposition to L}
    \mathcal{E}^{(2)(2b)}_{\ell} =  \sum_{L=0}^{2\ell} [\mathcal{E}^{(2)(2b)}_{\ell}]_{L0}
     \, ,
\end{equation}
actually contributes. Selecting such a scalar part, only one of the two projections on $J=0$ required to compute $\rho^{(2)}_{\text{QR}}(\mathbf{r}_{1}, \mathbf{r}_{2})$ is in fact necessary in Eq.~\eqref{eccPHFB}. While this result is trivial to prove based on the left-hand side of Eq.~\eqref{eccPHFB} by commuting the projector $P^{J=0\dagger}$ with $[\mathcal{E}^{(2)(2b)}_{\ell}]_{00}$ before using the hermiticity and idempotency of $P^{J=0}$, let us now sketch how such a result can be recovered from the right-hand side of Eq.~\eqref{eccPHFB}.

First, it is trivial to prove that
\begin{equation}
\rho^{(2)}_{\Omega'\Omega}(\mathbf{r}_{1},\mathbf{r}_{2}) = \rho^{(2)}_{0\Omega-\Omega'}({\cal R}^{-1}_{J}(\Omega')\mathbf{r}_{1},{\cal R}^{-1}_{J}(\Omega')\mathbf{r}_{2})
     \, ,
\end{equation}
where 
\begin{align}
a^{(\dagger)} ({\cal R}^{-1}_{J}(\Omega')\mathbf{r}_{i})   &\equiv R^{\dagger}_{J}(\Omega') a^{(\dagger)} (\mathbf{r}_{i}) R_{J}(\Omega') \, ,\\
R_{J}(\Omega-\Omega') &\equiv R^{\dagger}_{J}(\Omega')R_{J}(\Omega) \, .
\end{align}
Performing the changes of variables
\begin{align}
\mathbf{r'}_{i} &\equiv {\cal R}^{-1}_{J}(\Omega')\mathbf{r}_{i} \, ,\\
\Omega'' &\equiv \Omega-\Omega'\, ,
\end{align}
and exploiting the scalar nature of the operator translating into
\begin{equation}
[\mathcal{E}^{(2)(2b)}_{\ell}]_{00}({\cal R}_{J}(\Omega')\mathbf{r'}_{1},{\cal R}_{J}(\Omega')\mathbf{r'}_{2}) = [\mathcal{E}^{(2)(2b)}_{\ell}]_{00}(\mathbf{r'}_{1},\mathbf{r'}_{2})\, ,
\end{equation}
one eventually obtains that
\begin{align}
\langle \Theta^{0^+}_{\text{PHFB}}  | \mathcal{E}^{(2)(2b)}_{\ell}| \Theta^{0^+}_{\text{PHFB}}\rangle 
    &= \int_{\mathbf{r}_{1,2}} \!  \frac{[\mathcal{E}^{(2)(2b)}_{\ell}]_{00}(\mathbf{r}_{1},\mathbf{r}_{2})}{8\pi^2c^2_{0}} \int_{\Omega} \rho^{(2)}_{0\Omega}(\mathbf{r}_{1},\mathbf{r}_{2})  \, , \label{eccPHFB2}
\end{align}
such that a single integral over the Euler angles needs to be performed in practice.

\section{Quantum rigid rotor}

\subsection{Needle approximation}

The first step is to connect the QR model to the quantum {\it rigid} rotor (QRR) model by approximating the quantum symmetry restorations at play in PHFB through their classical counterpart. This is achieved via the so-called {\it needle approximation} that consists of considering that the overlap between two Bogoliubov states characterized by their orientations in the intrinsic frame is sharply peaked, i.e., it is non zero only if both states point strictly in the same direction such that 
\begin{equation}
 \langle \Phi(\Omega') | \Phi(\Omega,\varphi) \rangle \approx \delta(\Omega'-\Omega)\delta(\varphi) \, . \label{needleapprox}
\end{equation}

\subsection{One- and two-body densities}

Inserting Eq.~\eqref{needleapprox} into Eq.~\eqref{normalization} and Eqs.~\eqref{1bodyoffdiagPHFBdensity}-\eqref{2bodyoffdiagPHFBdensity}, leads to
\begin{subequations}
\label{PHFBdensityrigid}
\begin{align}
c^2_{0} \approx& \frac{1}{2\pi} \frac{1}{8\pi^2} \, , \label{normPHFBdensityrigid} \\
\rho^{(1)}_{\Omega'\Omega}(\mathbf{r}_{1}) \approx&  \frac{1}{2\pi} \rho^{(1)}_{\Omega\Omega0}(\mathbf{r}_{1},\mathbf{r}_{1})  \, \delta(\Omega'-\Omega) \label{1bodyPHFBdensityrigid}  \, , 
\\
\rho^{(2)}_{\Omega'\Omega}(\mathbf{r}_{1},\mathbf{r}_{2}) \approx&  \frac{1}{2\pi}  \rho^{(2)}_{\Omega\Omega0}(\mathbf{r}_{1},\mathbf{r}_{2})  \, \delta(\Omega'-\Omega) \label{2bodyPHFBdensityrigid}  \\
\approx& \frac{1}{2\pi} \left[\rho^{(1)}_{\Omega\Omega0}(\mathbf{r}_{1},\mathbf{r}_{1})\rho^{(1)}_{\Omega\Omega0}(\mathbf{r}_{2},\mathbf{r}_{2}) \right.  \nonumber \\
& -\rho^{(1)}_{\Omega\Omega0}(\mathbf{r}_{2},\mathbf{r}_{1})\rho^{(1)}_{\Omega\Omega0}(\mathbf{r}_{1},\mathbf{r}_{2}) \nonumber \\
&\left. + \bar{\kappa}^{(1)\ast}_{\Omega\Omega0}(\mathbf{r}_{1},\mathbf{r}_{2})\kappa^{(1)}_{\Omega\Omega0}(\mathbf{r}_{2},\mathbf{r}_{1}) \right]  \, \delta(\Omega'-\Omega) \, . \nonumber
\end{align}
\end{subequations}
Further inserting Eq.~\eqref{PHFBdensityrigid} into Eq.~\eqref{PHFBdens} delivers the QRR approximations $\rho^{(1)}_{\text{QRR}}(\mathbf{r}_{1})$ and $\rho^{(2)}_{\text{QRR}}(\mathbf{r}_{1},\mathbf{r}_{2})$ to the one and two-body PHFB local densities under the form given by Eq.~\eqref{SM_kbodydensityRR}. 

\subsection{Two-body mean-square eccentricity}

Inserting $\rho^{(2)}_{\text{QRR}}(\mathbf{r}_{1},\mathbf{r}_{2})$ into Eq.~\eqref{meansquareeccentricity} provides the QRR approximation to the two-body part of the normalized mean-square eccentricity. Alternatively, one can insert Eqs.~\eqref{normPHFBdensityrigid} and~\eqref{2bodyPHFBdensityrigid} into Eq.~\eqref{eccPHFB2} to obtain the approximated two-body part of the un-normalized mean-square eccentricity under the form
\begin{align}
\langle \Theta^{0^+}_{\text{PHFB}}  | \mathcal{E}^{(2)(2b)}_{\ell}| \Theta^{0^+}_{\text{PHFB}}\rangle_{r} 
    &= \int_{\mathbf{r}_{1,2}} \! [\mathcal{E}^{(2)(2b)}_{\ell}]_{00}(\mathbf{r}_{1},\mathbf{r}_{2}) \, \rho^{(2)}_{00}(\mathbf{r}_{1},\mathbf{r}_{2})  \, , \label{eccPHFB3}
\end{align}
which leads to realizing that the needle approximation delivers nothing but the symmetry-breaking HFB result. With this at hand, Eq.~\eqref{eq: SM Beta parameter definition1} is used to obtain the QRR approximation $\mathcal{B}^{2}_{\ell}(\text{HE})_{\text{QRR}}$ to the QR effective quadrupole deformation. 

\subsection{Needle approximation to a scalar operator}

The needle approximation to the PHFB expectation value of the two-body mean-square eccentricity was discussed just above. For the purpose of the analysis, let us here rewrite the needle approximation (i) for an arbitrary scalar $O$ such as the Hamiltonian $H$ or the squared total angular momentum $J^2$ while (ii) limiting oneself for simplicity to an axially deformed and parity conserving deformed HF state such that no projection on particle number or parity is necessary in the first place. 

The PHF expectation value can be trivially written as 
\begin{align}
\langle \Theta^{0^+}_{\text{PHF}}  | O | \Theta^{0^+}_{\text{PHF}}\rangle 
    &= \frac{\int_{\Omega} o(\Omega) \langle \Phi | \Phi(\Omega) \rangle}{\int_{\Omega} \langle \Phi | \Phi(\Omega) \rangle}  \, , \label{expPHF}
\end{align}
where the so-called reduced operator kernel $o(\Omega)$ amenable to the application of the off-diagonal Wick's theorem is given by
\begin{align}
o(\Omega) &\equiv  \frac{\langle \Phi | O | \Phi(\Omega) \rangle}{\langle \Phi | \Phi(\Omega) \rangle} \, . \label{Oredkernel}
\end{align}
As discussed previously, the needle approximation characterized by $\langle \Phi | \Phi(\Omega) \rangle \approx \delta(\Omega)$ delivers 
\begin{align}
\langle \Theta^{0^+}_{\text{PHF}}  | O | \Theta^{0^+}_{\text{PHF}}\rangle_{r} 
    &= \langle \Phi | O | \Phi \rangle  \, , \label{expPHFr}
\end{align}
which is nothing but the symmetry breaking HF expectation value.  While the validity of the needle approximation, i.e., the extent to which the quantum rigid rotor  reproduces the pure quantum rotor, primarily depends how peaked around zero the overlap kernel $\langle \Phi | \Phi(\Omega) \rangle$ is, Eq.~\eqref{expPHF} also makes clear that the quality of the approximation cannot be stated in general, i.e., it depends on the operator $O$ and more specifically on the behavior of the operator kernel $o(\Omega)$ as a function of $\Omega$.

\section{Classical rigid rotor}
\label{CRRagain}

Equation~\eqref{2bodyPHFBdensityrigid} allows one to decompose the QRR two-body local density into the so-called {\it direct} (d), {\it exchange} (e) and {\it pairing (p)} contributions
\begin{align}
\rho^{(2)}_{\text{QRR}}(\mathbf{r}_{1},\mathbf{r}_{2}) &\equiv \sum_{x=d,e,p} \rho^{(2)}_{\text{QRR}_{x}}(\mathbf{r}_{1},\mathbf{r}_{2}) \, . \label{decompositionrho2}
\end{align}
Inserting Eq.~\eqref{decompositionrho2} into Eq.~\eqref{eq: SM Beta parameter definition1} leads to identifying three distinct contributions to $\mathcal{B}^{2}_{2}(\text{HE})_{\text{QRR}}$ according to
\begin{equation}
\mathcal{B}^{2}_{2}(\text{HE})_{\text{QRR}} \equiv \sum_{x=d,e,p} \mathcal{B}^{2}_{2}(\text{HE})_{\text{QRR}_{x}} \, .
\end{equation}

Equation~\eqref{decompositionrho2} makes clear that approximating the quantum symmetry projection at play in the QR via the needled approximation does not, by itself, deliver the CRR given that purely quantal two-body correlations associated with Pauli's exclusion principle and superfluidity, i.e., the fermionic character of nucleons, are still included in the description. Reaching the CRR description requires to further omit the exchange and pairing contributions to the two-body local density in Eq.~\eqref{decompositionrho2} such that no correlations at all are present in the intrinsic frame. Only retaining the direct contribution, one eventually obtains
\begin{align}
\rho^{(2)}_{\text{QR}}(\mathbf{r}_{1},\mathbf{r}_{2}) &\approx \rho^{(2)}_{\text{QRR}_{d}}(\mathbf{r}_{1},\mathbf{r}_{2}) \nonumber \\
&=\frac{1}{8\pi^2} \int_{\Omega} d\Omega \, \rho^{(1)}_{\Omega\Omega}(\mathbf{r}_{1}) \rho^{(1)}_{\Omega\Omega}(\mathbf{r}_{2}) \nonumber \\
&\equiv \rho^{(2)}_{\text{CRR}}(\mathbf{r}_{1},\mathbf{r}_{2}) \, . \label{RRapproxtoPHFBdens}
\end{align}
such that\footnote{Starting from Eq.~\eqref{eccPHFB3}, it is clear that the (un-normalized) two-body part of the mean-square eccentricity is now approximated as
$\displaystyle \protect\langle \protect\Theta^{0^+}_{\text{PHFB}}  | \protect\mathcal{E}^{(2)(2b)}_{\protect\ell}| \protect\Theta^{0^+}_{\protect\text{PHFB}}\protect\rangle_{d}
    = \protect\int_{\protect\mathbf{r}_{1,2}} \protect\! [\protect\mathcal{E}^{(2)(2b)}_{\protect\ell}]_{00}(\protect\mathbf{r}_{1},\protect\mathbf{r}_{2}) \, \protect\rho^{(1)}_{00}(\protect\mathbf{r}_{1})\protect\rho^{(1)}_{00}(\protect\mathbf{r}_{2})  \protect\, , \protect\nonumber$
which is nothing but the direct part of the symmetry-breaking HFB contribution.}
\begin{align}
\mathcal{B}^{2}_{2}(\text{HE})_{\text{QRR}_{d}}& \equiv \mathcal{B}^{2}_{2}(\text{HE})_{\text{CRR}} \, . \label{RRapproxtoPHFBdef}
\end{align}

\section{Ab initio results}

The present analysis relies first on PHFB calculations based on the EM1.8/2.0~\cite{Hebeler11a} Chiral effective field theory ($\chi$EFT) Hamiltonian. Operators are expanded using a one-body spherical harmonic oscillator basis (sHO) characterized by the frequency  $\hbar\omega = 12 $\,MeV and truncated to 11 major shells, i.e., $e_\mathrm{1max} = 10$. Three-body matrix elements are limited to three-body basis states characterized by $e_\mathrm{3max} = 24<3e_\mathrm{1max}$ and the three-nucleon interaction is approximated as an effective two-nucleon interaction through a rank-reduction method~\cite{Frosini21a}.

The body of the paper focuses on PHFB results of five representative even-even nuclei: $^{20}\mathrm{Ne}$, $^{28}\mathrm{Si}$, $^{40}\mathrm{Ca}$, $^{76}\mathrm{Ge}$ and $^{136}\mathrm{Xe}$. The projections on angular momentum, as well as on neutron and proton numbers, are performed on top of axially deformed HFB states constrained to display a large range of intrinsic quadrupole deformations $\beta_{20}[\text{dHFB}]$.

The present section extends the discussion provided in the main body of the paper to a large set of even-even nuclei ground-states with  $8\leq Z \leq 28$~\cite{Bofos:2026huw}. The projections on angular momentum, as well as on neutron and proton numbers, are performed on top of the deformed axial HFB minimum characterized by $\beta_{20}[\text{dHFB}]$.

\subsection{From the quantum rotor to the quantum rigid rotor}

\begin{figure}[htbp]     
  \centering             
  \includegraphics[width=0.4\paperwidth]{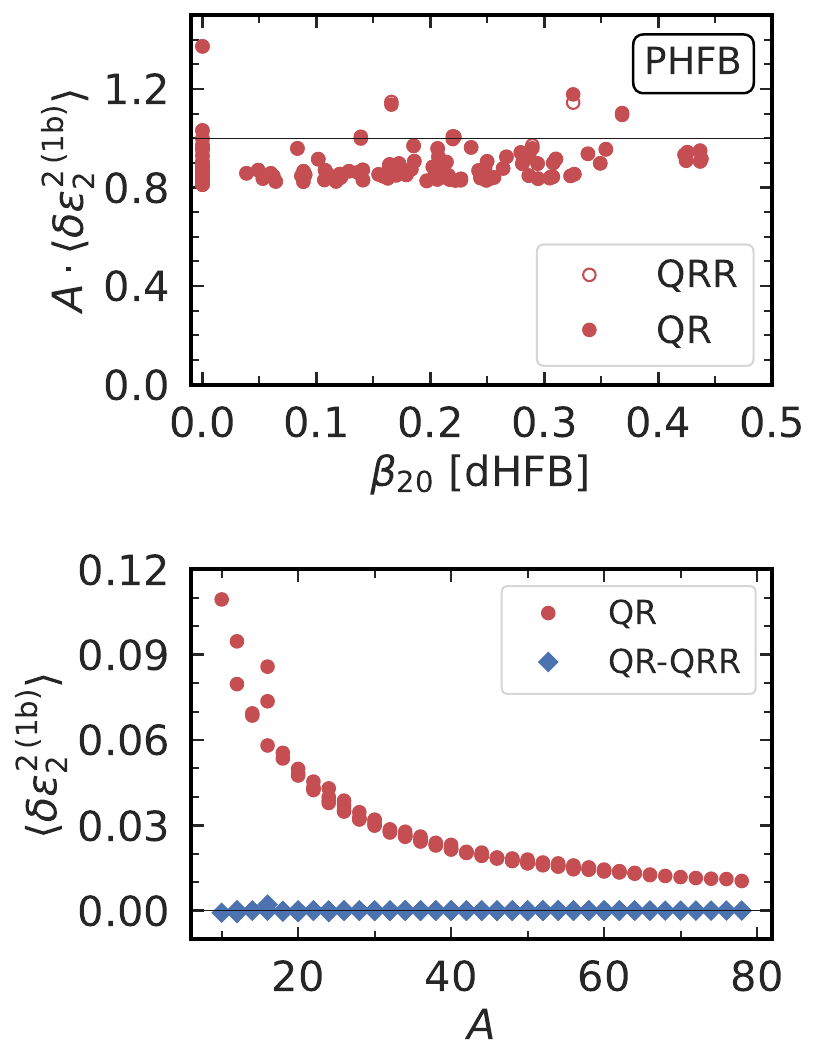}
  \caption{One-body contribution to the {\it ab initio} normalized quadrupole mean-square eccentricity computed with the quantum rotor (QR) and the quantum rigid rotor (QRR) approximations for a large set of even-even nuclei ground-states with  $8\leq Z \leq 28$. Upper panel: QR and QRR values as a function of $\beta^2_{20}[\text{dHFB}]$. Lower panel: QR values as well as difference between QR and QRR values as a function of $A$ .}
  \label{fig:ecc1b_diff_rigid_quant_A}
\end{figure}

The impact of the needle, i.e., rigid, approximation to the fully quantum symmetry restoration on the one-body contribution to the normalized quadrupole mean-square eccentricity is quantified in Fig.~\ref{fig:ecc1b_diff_rigid_quant_A} as a function of both  $\beta^2_{20}[\text{dHFB}]$ and $A$. The negligible difference between the QR and its QRR approximation displayed in the bottom panel demonstrates that the needle approximation has zero impact on this one-body contribution. Further realizing that  for a one-body operator the QRR is equal to the CRR approximation, this result clarifies why QR results are already consistent with the CRR approximation stipulating that the one-body contribution to the normalized quadrupole mean-square eccentricity is independent of the intrinsic deformation and scales with $A^{-1}$ ~\cite{Bofos:2026huw}.

\begin{figure}[htbp]     
  \centering             
  \includegraphics[width=0.4\paperwidth]{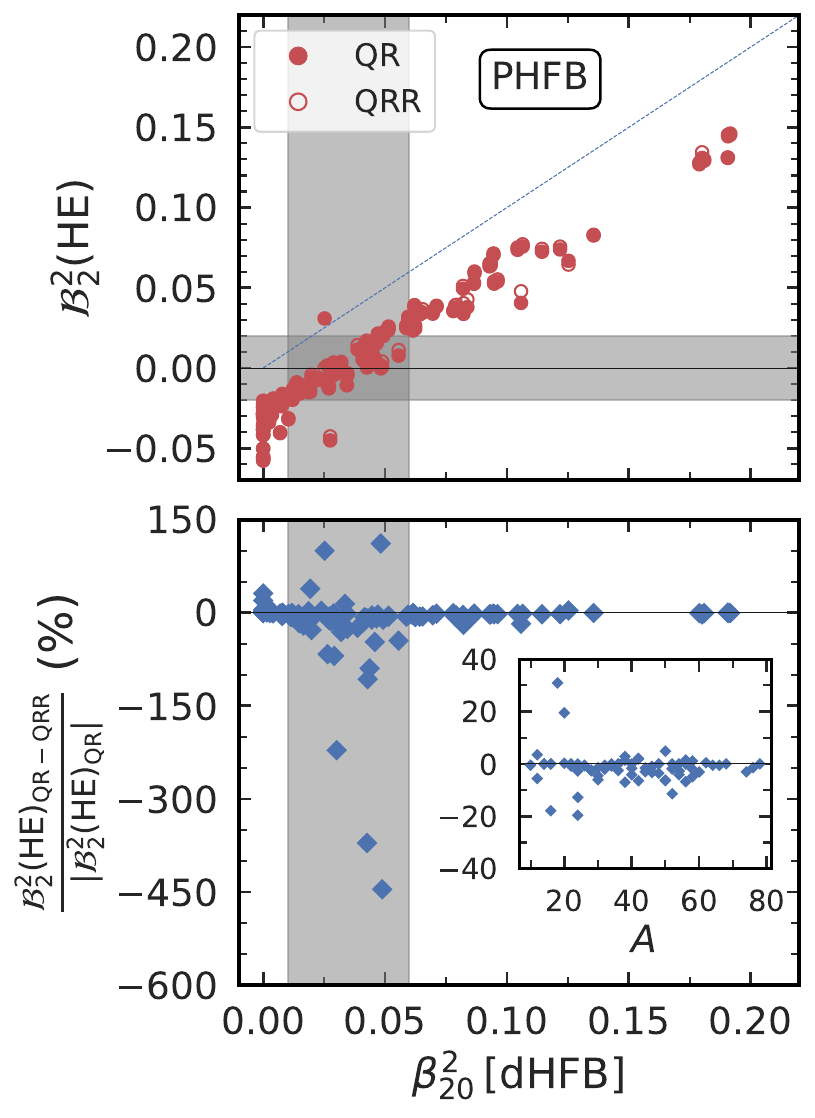}
  \caption{{\it Ab initio} effective deformation ${\cal B}_2^2(\mathrm{HE})$ against $\beta^2_{20}[\text{dHFB}]$ for a large set of even-even nuclei ground-states with  $8\leq Z \leq 28$ computed within the quantum rotor (QR) and the quantum rigid rotor (QRR) approximations. Upper panel: absolute values. The dashed line shows the first diagonal. Lower panel: relative error of the QRR with respect to the QR. The grey bands highlight  nuclei whose effective deformation is close to 0, i.e., for which ${\cal B}_2^2(\mathrm{HE})_{\text{QR}} \in [-0.02,0.02]$. The inset displays the relative error as a function of $A$ while excluding the points within the grey bands.}
  \label{fig:B2sq_diff_rigid_quant_beta2}
\end{figure}

Moving to the two-body contribution to the normalized quadrupole mean-square eccentricity, the top panel of Fig.~\ref{fig:B2sq_diff_rigid_quant_beta2} displays the effective quadrupole deformation ${\cal B}_2^2(\mathrm{HE})$ as a function of $\beta^2_{20}[\text{dHFB}]$. As discussed in Ref.~\cite{Bofos:2026huw}, the QR effective deformation is strongly correlated with $\beta^2_{20}[\text{dHFB}]$ but it is not equal to it and it is even negative for $\beta^2_{20}[\text{dHFB}]\leq 0.03$. Comparing the QR and its QRR approximation, the needle approximation to the quantum projection is seen to deliver a very good approximation, i.e., the absolute error induced on the effective deformation is indeed very small. 

As visible from the bottom panel of Fig.~\ref{fig:B2sq_diff_rigid_quant_beta2}, the {\it relative} error on the effective deformation is however not negligible whenever ${\cal B}_2^2(\mathrm{HE})_{\text{QR}}$ is very small, i.e., the relative error can reach $-450\%$ whenever ${\cal B}_2^2(\mathrm{HE})_{\text{QR}} \in [-0.02,0.02]$, which happens for $\beta^2_{20}[\text{dHFB}] \in [0.01,0.06]$ (i.e., $|\beta_{20}[\text{dHFB}]| \in [0.1,0.25]$). One must remember that this smallness of the effective deformation parameter is somewhat fictitious given that additional, i.e., vibrational and non-collective, correlations non included in the present study will eventually contribute to push ${\cal B}_2^2(\mathrm{HE})$ up away from zero. Once this is taken into account, the relative error associated with the classical approximation to the symmetry restoration will be strongly suppressed. Anticipating this aspect, the inset in the bottom panel  of Fig.~\ref{fig:B2sq_diff_rigid_quant_beta2} shows the relative error as a function of $A$ while excluding the points corresponding to ${\cal B}_2^2(\mathrm{HE})_{\text{QR}} \in [-0.02,0.02]$. While the error is significant (up to $30\%$) for a few light systems, it is typically smaller than $10\%$ for the large majority of nuclei. This is consistent with the analysis conducted in Ref.~\cite{Ke:2025}.

\section{MR-EDF results}

The present analysis further relies on PHFB calculations performed within the MR-EDF framework based on the PC-F1  covariant EDF parameterization~\cite{Burvenich:2002}. Operators are expanded using a one-body sHO basis characterized by the frequency  $\hbar\omega = 41 A^{-1/3}$ MeV and truncated according to $e_{\rm 1max}=8$ for $A   \le 50$, $e_{\rm 1max}=10$ for $50 < A \le 100$, $e_{\rm 1max}=12$ for $100 < A \le 150$, and $e_{\rm 1max}=14$ for $A > 150$.

The discussion provided in the main-body of the text relates to systematic PHFB results  obtained for a very large range of even-even nuclei with $Z \in [4,100]$. The projections on angular momentum, as well as on neutron and proton numbers, are performed on top of the deformed axial HFB minimum characterized by  $\beta_{20}[\text{dHFB}]$. 

The analysis is presently complemented by focusing on four representative nuclei ($^{20}$Ne, $^{76}$Ge, $^{136}$Xe, $^{224}$Ra) while placing a constraint on the quadrupole intrinsic deformation $Q_{20}[\text{dHFB}]$ of the underlying HFB state.

\subsection{From the quantum rotor to the quantum rigid rotor}


The impact of the needle approximation on the one-body contribution to the normalized mean-square eccentricity is not explicitly reported here given that it is fully consistent with the one reported in Fig.~\ref{fig:ecc1b_diff_rigid_quant_A} for {\it ab initio} calculations: it  is negligible.

\begin{figure}[htbp]     
  \centering             
  \includegraphics[width=0.44\textwidth]{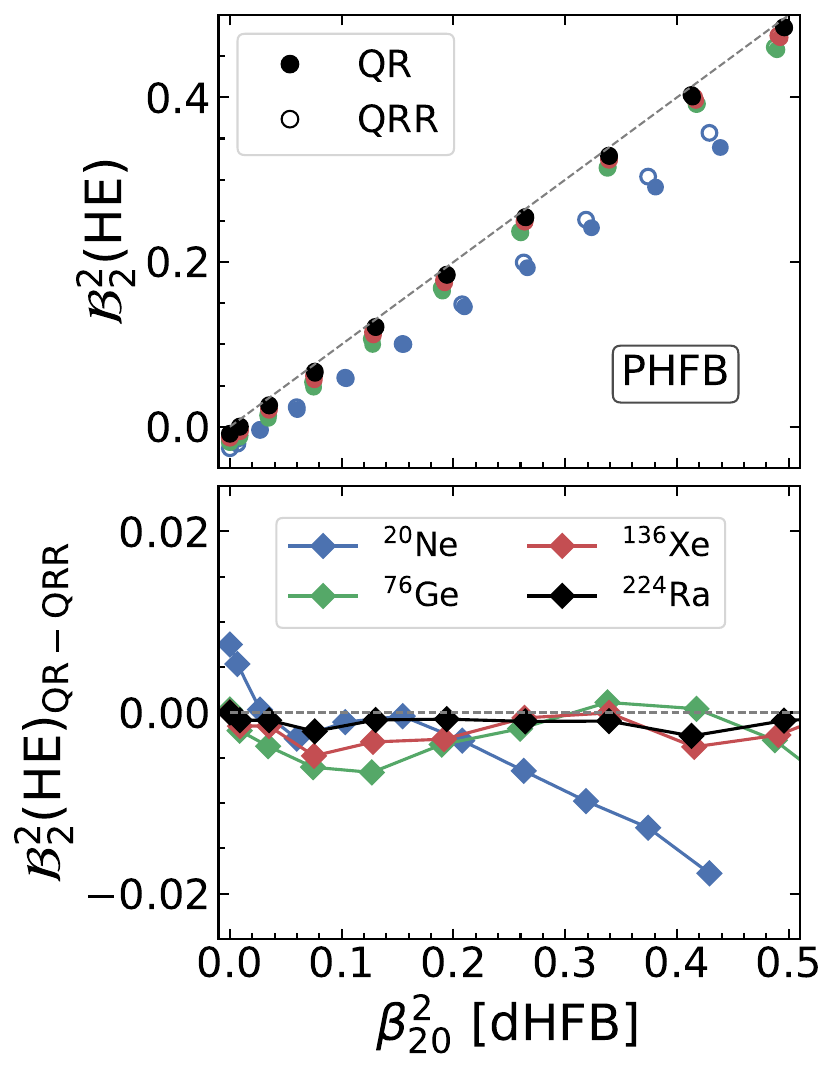}
  \caption{MR-EDF effective deformation ${\cal B}_2^2(\mathrm{HE})$ against $\beta^2_{20}[\text{dHFB}]$ for $^{20}\mathrm{Ne}$, $^{76}\mathrm{Ge}$, $^{136}\mathrm{Xe}$ and $^{224}\mathrm{Ra}$ computed within the quantum rotor (QR) and the quantum rigid rotor (QRR) approximations. Upper panel: absolute values. The dashed line shows the first diagonal. Lower panel: absolute error of the QR with respect to the QRR.}
  \label{FigMREDFSM2}
\end{figure}

The top panel of Fig.~\ref{FigMREDFSM2} displays the QR and QRR effective quadrupole deformations ${\cal B}_2^2(\mathrm{HE})$ as a function of $\beta^2_{20}[\text{dHFB}]$ for  $^{20}\mathrm{Ne}$, $^{76}\mathrm{Ge}$, $^{136}\mathrm{Xe}$ and $^{224}\mathrm{Ra}$. Consistently with the {\it ab initio} results discussed in Ref.~\cite{Bofos:2026huw} and in the present work, the QR effective deformation is strongly correlated with $\beta^2_{20}[\text{dHFB}]$ without being equal to it. The offset with respect to the first diagonal quickly decreases with $A$. 

The comparison between the QR effective deformation and its QRR counterpart proves that the absolute error induced by the needle approximation is very small. As the bottom panel illustrates, this small error however evolves non-trivially with both $A$ and $\beta^2_{20}[\text{dHFB}]$, e.g.,it is the largest in $^{20}$Ne where it also changes sign with increasing deformation.


\begin{figure}[htbp]     
  \centering             
  \includegraphics[width=0.4\paperwidth]{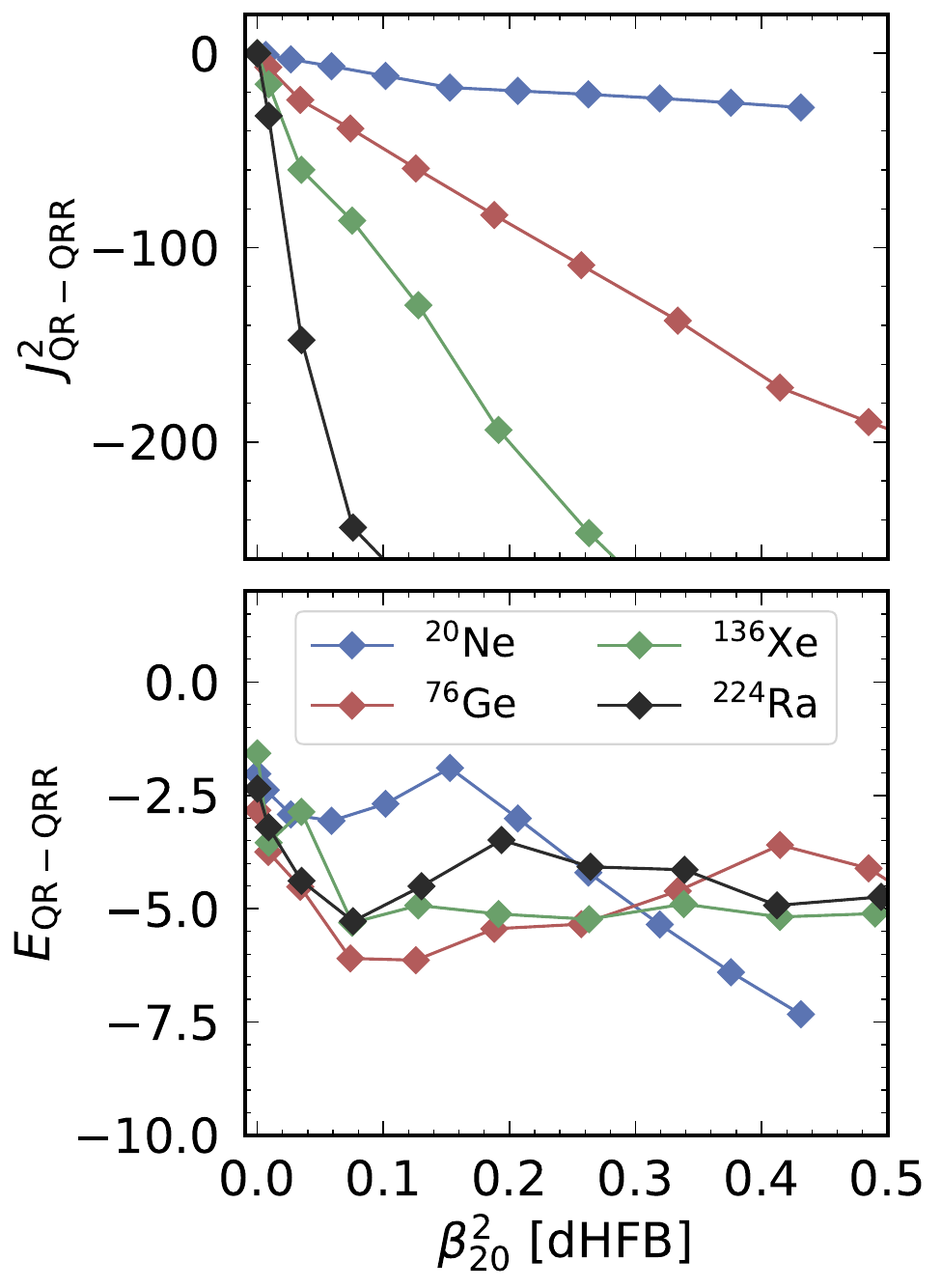}
  \caption{Error on $J^2$ (upper panel) and on the energy (lower panel) as a function of $\beta^2_{20}[\text{dHFB}]$ due to the needle approximation to the symmetry restoration in MR-EDF calculations of $^{20}\mathrm{Ne}$, $^{76}\mathrm{Ge}$, $^{136}\mathrm{Xe}$ and $^{224}\mathrm{Ra}$.}
  \label{FigMREDFSM3}
\end{figure}

As a matter of fact, the validity of the needle approximation, i.e.,the extent to which the QRR reproduces the QR, cannot be stated in general. The size of the error and its evolution with, e.g., $A$ and $\beta^2_{20}[\text{dHFB}]$ depends on the observable $O$ under consideration. As Eq.~\eqref{expPHF} makes clear, the quality of the approximation reflects the dependence on $\Omega$ of both the norm kernel $\langle \Phi | \Phi(\Omega) \rangle$ and the operator kernel $o(\Omega)$. To illustrate this point, Fig.~\ref{FigMREDFSM3} further displays  as a function of $\beta^2_{20}[\text{dHFB}]$ the error induced by the needle approximation on $J^2$ (whose QR value is zero for all $\beta_{20}[\text{dHFB}]$) and on the total energy. The negative error on $J^2$ grows  significantly and monotonically with both $A$ and $\beta^2_{20}[\text{dHFB}]$. As for the energy, the negative error remains modest. While it slightly increases at first, it quickly saturates with $A$. Similarly this error first increases with deformation before decreasing beyond $\beta^2_{20}[\text{dHFB}]\sim 0.2-0.3$ ($\beta_{20}[\text{dHFB}]\sim 0.4-0.6$).



\subsection{From the quantum rigid rotor to the classical rigid rotor}

\begin{figure}[htbp]     
  \centering             
  \includegraphics[width=0.44\textwidth]{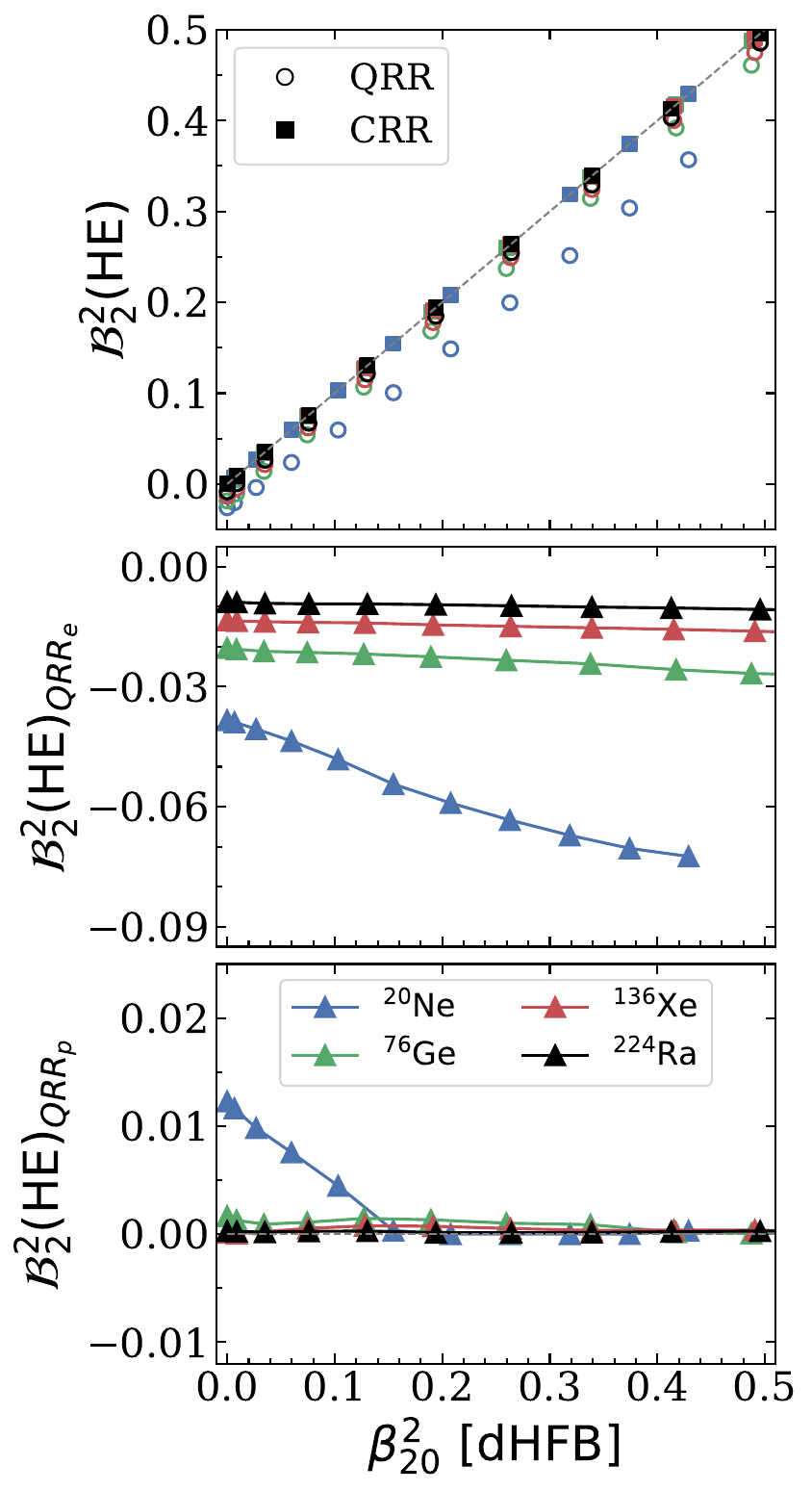}
  \caption{MR-EDF effective deformation ${\cal B}_2^2(\mathrm{HE})$ against $\beta^2_{20}[\text{dHFB}]$ for $^{20}\mathrm{Ne}$, $^{76}\mathrm{Ge}$, $^{136}\mathrm{Xe}$ and $^{224}\mathrm{Ra}$. Upper panel: QRR and CRR values. The dashed line shows the first diagonal. Middle panel: contribution from the exchange term to the QRR effective deformation. Bottom panel: contribution from the pairing term to the QRR effective deformation.  
  }
  \label{fig:MR_CDFT_Fig3}
\end{figure}

Figure~\ref{fig:MR_CDFT_Fig3} extends the discussion conducted in the main body of the text  within the {\it ab initio} framework to PHFB calculations performed within the MR-EDF framework. The upper panel compares the effective deformations ${\cal B}_2^2(\mathrm{HE})$ computed within the QRR and CRR approximations as a function of  $\beta^2_{20}[\text{dHFB}]$ for $^{20}\mathrm{Ne}$, $^{76}\mathrm{Ge}$, $^{136}\mathrm{Xe}$ and $^{224}\mathrm{Ra}$. As explained in the formal section above, the difference relates to the pure quantum contributions to the two-body part of the QRR mean-square eccentricity associated with the so-called exchange and pairing terms. The present results confirm that removing both contributions brings the effective deformation back on the diagonal, thus recovering the results derived analytically in the formal section above and stipulating that $B_2^2(\mathrm{HE})_{\text{CRR}}=\beta^2_{20}[\text{dHFB}]$. The middle and bottom panels single out the two subtracted contributions, thus showing that the exchange term largely dominates. Furthermore, while the exchange term grows linearly with $\beta^2_{20}[\text{dHFB}]$, the pairing term quickly decreases with it. Last but not least, one observes that both terms decrease monotonically with $A$ and would indeed converge to zero in the $A\rightarrow \infty$ limit.

\begin{figure}[htbp]     
  \centering             
  \includegraphics[width=0.4\paperwidth]{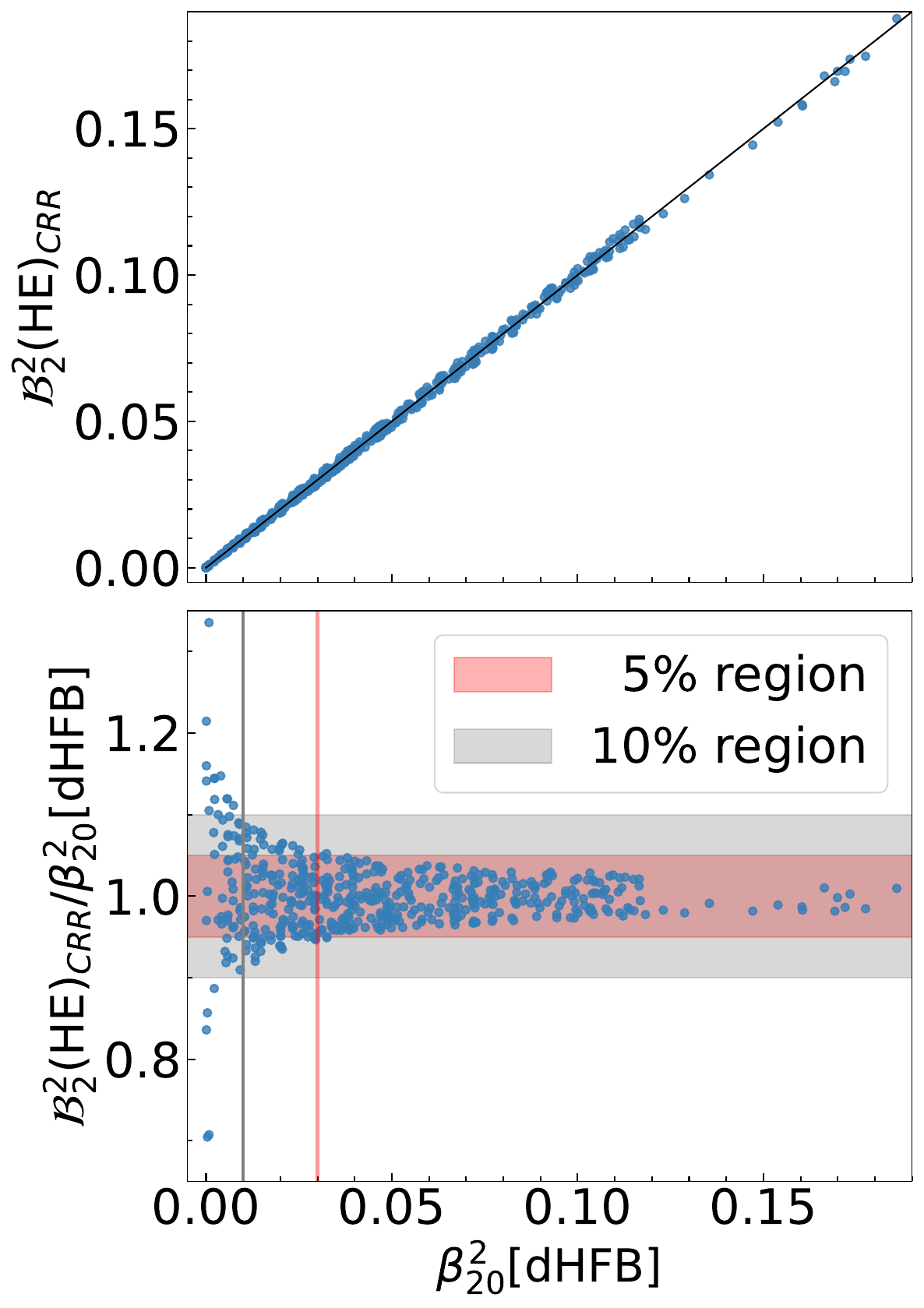}
  \caption{Comparison of the MR-EDF effective quadrupole deformation computed within the CRR approximation with the intrinsic axial quadrupole deformation $\beta^2_{20}[\text{dHFB}]$. Upper panel: ${\cal B}_2^2(\mathrm{HE})_{\text{CRR}}$ against $\beta^2_{20}[\text{dHFB}]$. Bottom panel: ratio ${\cal B}_2^2(\mathrm{HE})_{\text{CRR}}/\beta^2_{20}[\text{dHFB}]$ as a function of $\beta^2_{20}[\text{dHFB}]$. }
  \label{FigMREDFSM4}
\end{figure}

\begin{figure}[htbp]     
  \centering             
  \includegraphics[width=0.4\paperwidth]{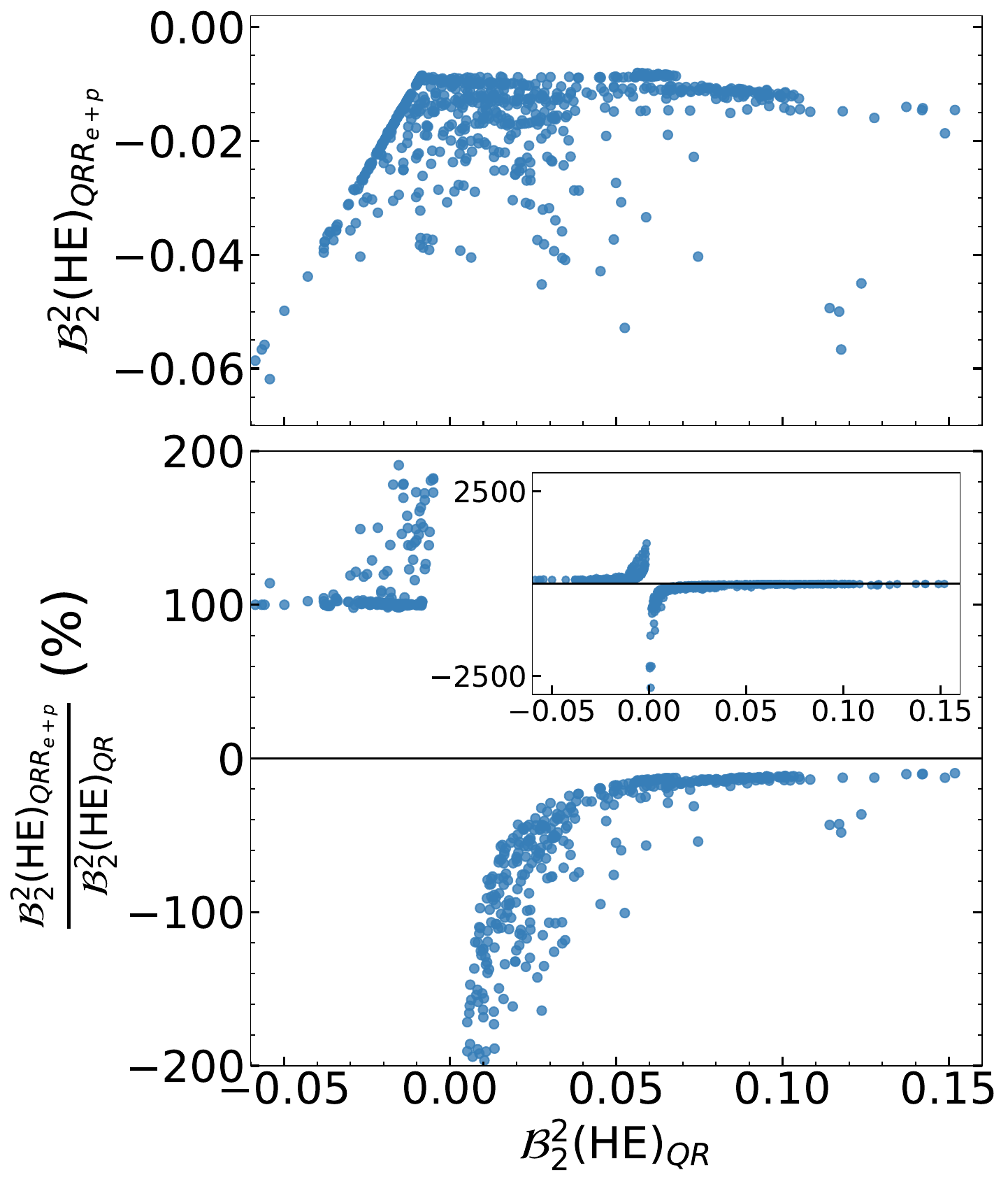}
  \caption{Pure quantum contributions $\mathcal{B}^{2}_{2}(\text{HE})_{\text{QRR}_{e+p}}$ to the MR-EDF effective deformation within the QR approximation  $\mathcal{B}^{2}_{2}(\text{HE})_{\text{QR}}$ against the latter. Upper panel: $\mathcal{B}^{2}_{2}(\text{HE})_{\text{QRR}_{e+p}}$. Bottom panel: $\mathcal{B}^{2}_{2}(\text{HE})_{\text{QRR}_{e+p}}/\mathcal{B}^{2}_{2}(\text{HE})_{\text{QR}}$. While the vertical range is limited to ratios smaller than $200\%$ (in absolute) whereas the inset displays the same information up to ratios equal to 2500\%.}
  \label{FigMREDFSM5}
\end{figure}

Figures~\ref{FigMREDFSM4} and \ref{FigMREDFSM5} refine the above analysis by detailing two key aspects of the systematic MR-EDF results. First, Fig.~\ref{FigMREDFSM4} quantifies to which extent the rigid plus classical approximations to the QR effective quadrupole deformation do indeed deliver the intrinsic quadrupole deformation $\beta^2_{20}[\text{dHFB}]$. While the upper panel recalls the strong correlation between both quantities, the bottom panel stipulates that they do agree within $5\%$ as soon as $\beta^2_{20}[\text{dHFB}] \geq 0.03$ ($|\beta_{20}[\text{dHFB}]| \geq 0.17$), the agreement further improving monotonically with $\beta^2_{20}[\text{dHFB}]$.

Second, Fig.~\ref{FigMREDFSM5} quantifies systematically the pure quantum contribution $\mathcal{B}^{2}_{2}(\text{HE})_{\text{QRR}_{e+b}}$ associated with the fermionic character of nucleons to the effective deformation computed within the quantum rotor, i.e., PHFB, approximation. The upper panel confirms that this pure quantum contribution is always negative and that its absolute value remains below $0.06$. The bottom panel makes clear that, whenever the total QR effective deformation $\mathcal{B}^{2}_{2}(\text{HE})_{\text{QR}}$ is negative, it is entirely due to $\mathcal{B}^{2}_{2}(\text{HE})_{\text{QRR}_{e+b}}$. Whenever $\mathcal{B}^{2}_{2}(\text{HE})_{\text{QR}}$ is positive, the relative contribution of the pure quantum contribution is very large at first due to the overall smallness of $\mathcal{B}^{2}_{2}(\text{HE})_{\text{QR}}$ but decreases as a function of $\mathcal{B}^{2}_{2}(\text{HE})_{\text{QR}}$ to contribute for less than $10\%$ of it whenever $\mathcal{B}^{2}_{2}(\text{HE})_{\text{QR}}$ becomes larger than $0.14$.

\begin{figure*}[htbp]     
  \centering             
  \includegraphics[width=0.62\textwidth]{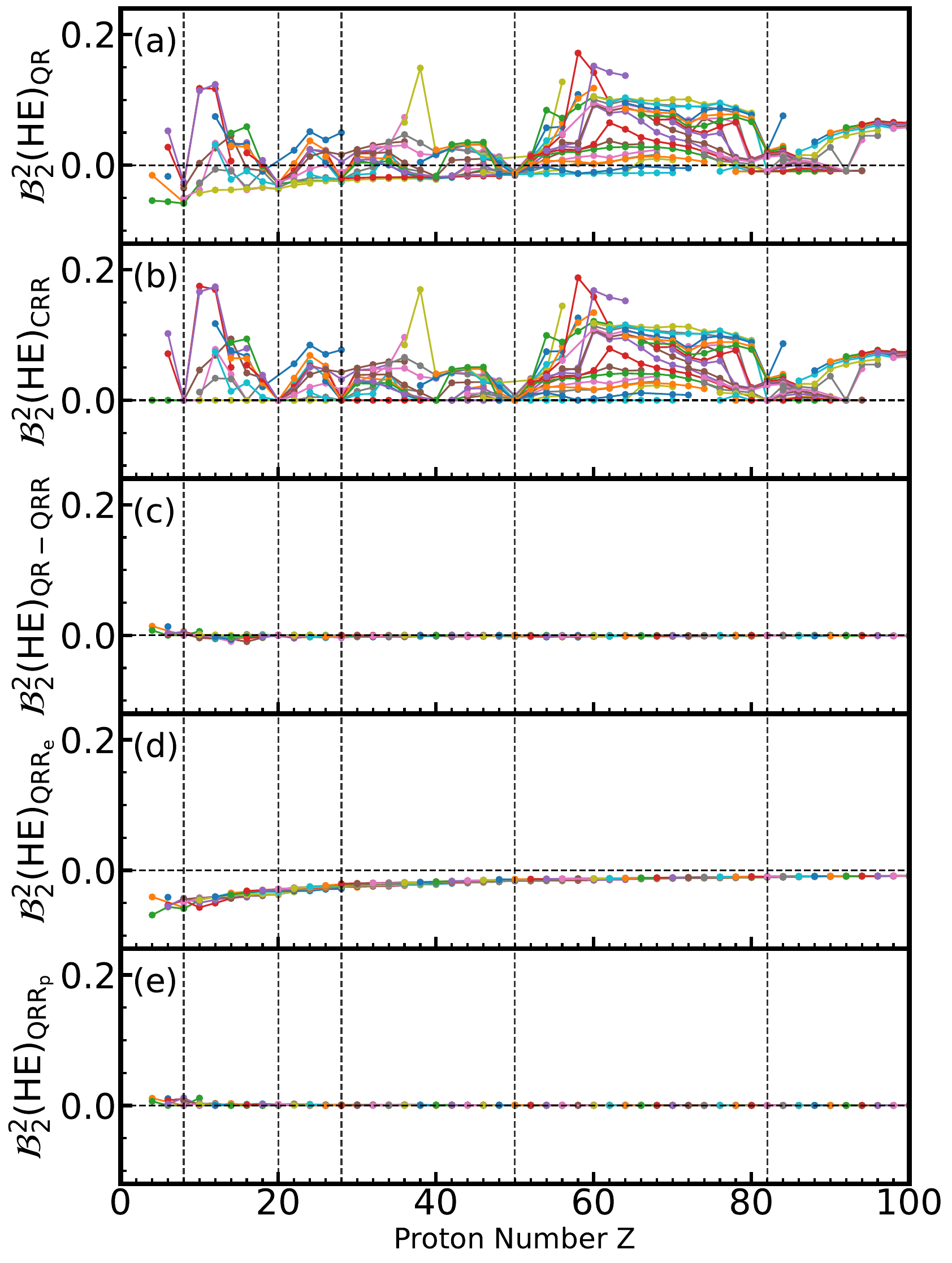}
  \caption{Systematic of MR-EDF effective deformation along many isotonic chains with $Z \in [4,100]$ as a function of proton number. (a) Effective deformation within the quantum rotor approximation. (b) Effective deformation within the classical rigid rotor approximation to the quantum rotor. 
  (c)  Absolute error of the quantum rotor approximation with respect to the quantum rigid rotor approximation.
  (d) Pure quantum contribution to the effective deformation within the quantum rigid rotor associated with the exchange term. (e) Pure quantum contribution to the effective deformation within the quantum rigid rotor associated with the pairing term. }
  \label{FigMREDFSM6}
\end{figure*}

Figure~\ref{FigMREDFSM6} provides the same systematic analysis as in the main body of the text, except that the theoretical data along isotonic chains are plotted  as a function of the proton number $Z$ rather than along isotopic chains as a function of the neutron number $N$. The conclusions of the analysis are essentially the same.

\bibliographystyle{apsrev4-1}   
\bibliography{ref.bib}

@article{Mehrabpour:2026yuc,
    author = "Mehrabpour, Hadi and Giacalone, Giuliano and Luzum, Matthew W.",
    title = "{Triaxial shapes and the angular structure of nuclear three-body correlations}",
    eprint = "2604.00619",
    archivePrefix = "arXiv",
    primaryClass = "nucl-th",
    month = "4",
    year = "2026"
}

@misc{lightions2025, 
      author={{CERN workshop}},
      title={Light-ion collisions at the {LHC} 2025},
      year={{Dec. 1–3, 2025, Geneva, Switzerland}}
}

@unpublished{frosini26a,
	Author = {Frosini, M. and Bally, B. and Duguet, T. and Scalesi, A. and Som\`a, V},
	Note = {Unpublished},
	Year = {2026}}

@article{Yao:2010,
    author = "Yao, J. M. and Meng, J. and Ring, P. and Vretenar, D.",
    title = "{Configuration mixing of angular-momentum projected triaxial relativistic mean-field wave functions}",
    eprint = "0912.2650",
    archivePrefix = "arXiv",
    primaryClass = "nucl-th",
    doi = "10.1103/PhysRevC.81.044311",
    journal = "Phys. Rev. C",
    volume = "81",
    pages = "044311",
    year = "2010"
}

@article{Yao:2019rck,
    author = "Yao, J. M. and Bally, B. and Engel, J. and Wirth, R. and Rodr\'\i{}guez, T. R. and Hergert, H.",
    title = "{$Ab Initio$ Treatment of Collective Correlations and the Neutrinoless Double Beta Decay of $^{48}$Ca}",
    eprint = "1908.05424",
    archivePrefix = "arXiv",
    primaryClass = "nucl-th",
    doi = "10.1103/PhysRevLett.124.232501",
    journal = "Phys. Rev. Lett.",
    volume = "124",
    number = "23",
    pages = "232501",
    year = "2020"
}

@article{Frosini2022a,
	author = {{Frosini, M.} and {Duguet, T.} and {Ebran, J.-P.} and {Som\`a, V.}},
	title = {Multi-reference many-body perturbation theory for nuclei - I. Novel PGCM-PT formalism},
	DOI= "10.1140/epja/s10050-022-00692-z",
	url= "https://doi.org/10.1140/epja/s10050-022-00692-z",
	journal = {Eur. Phys. J. A},
	year = 2022,
	volume = 58,
	number = 4,
	pages = "62",
}

@article{Frosini2022b,
	author = {{Frosini, M.} and {Duguet, T.} and {Ebran, J.-P.} and {Bally, B.} and {Mongelli, T.} and {Rodr\'{\i}guez, T. R.} and {Roth, R.} and {Som\`a, V.}},
	title = {Multi-reference many-body perturbation theory for nuclei - II. Ab initio study of neon isotopes via PGCM and IM-NCSM calculations},
	DOI= "10.1140/epja/s10050-022-00693-y",
	url= "https://doi.org/10.1140/epja/s10050-022-00693-y",
	journal = {Eur. Phys. J. A},
	year = 2022,
	volume = 58,
	number = 4,
	pages = "63",
}

@article{Sousa:2024msh,
    author = "Sousa, Jefferson and Noronha, Jorge and Luzum, Matthew",
    title = "{Initial energy-momentum to final flow: A general framework for heavy-ion collisions}",
    doi = "10.1103/PhysRevC.110.044909",
    journal = "Phys. Rev. C",
    volume = "110",
    number = "4",
    pages = "044909",
    year = "2024"
}

@article{Blaizot:2025scr,
    author = "Blaizot, Jean-Paul and Giacalone, Giuliano",
    title = "{Angular structure of many-body correlations in atomic nuclei}",
    eprint = "2504.15421",
    archivePrefix = "arXiv",
    primaryClass = "nucl-th",
    reportNumber = "CERN-TH-2025-057",
    doi = "10.1140/epja/s10050-025-01679-2",
    journal = "Eur. Phys. J. A",
    volume = "61",
    number = "9",
    pages = "220",
    year = "2025"
}

@article{Duguet2025a,
  title = {Revealing the Harmonic Structure of Nuclear Two-Body Correlations in High-Energy Heavy-Ion Collisions},
  author = {Duguet, Thomas and Giacalone, Giuliano and Jeon, Sangyong and Tichai, Alexander},
  journal = {Phys. Rev. Lett.},
  volume = {135},
  issue = {18},
  pages = {182301},
  numpages = {8},
  year = {2025},
  month = {Oct},
  publisher = {American Physical Society},
  doi = {10.1103/v2z7-wlnr},
  url = {https://link.aps.org/doi/10.1103/v2z7-wlnr}
}

@misc{Blaizot2025a, 
      author={Jean-Paul Blaizot and Giuliano Giacalone and Alessandro Lovato},
      year={2025},
      eprint={2512.18926},
      archivePrefix={arXiv},
      primaryClass={nucl-th},
      url={https://arxiv.org/abs/2512.18926}, 
}

@article{Duguet:2025qxi,
    author = "Duguet, T. and Giacalone, G. and Som{\`a}, V. and Zhou, Y.",
    title = "{Topical issue on the intersection of low-energy nuclear structure and high-energy nuclear collisions}",
    eprint = "2512.05874",
    archivePrefix = "arXiv",
    primaryClass = "nucl-th",
    doi = "10.1140/epja/s10050-025-01715-1",
    journal = "Eur. Phys. J. A",
    volume = "61",
    pages = "237",
    year = "2025"
}

@article{Giacalone2025a,
  title = {Exploiting $^{20}\mathrm{Ne}$ Isotopes for Precision Characterizations of Collectivity in Small Systems},
  author = {Giacalone, Giuliano and Bally, Benjamin and Nijs, Govert and Shen, Shihang and Duguet, Thomas and Ebran, Jean-Paul and Elhatisari, Serdar and Frosini, Mikael and L\"ahde, Timo A. and Lee, Dean and Lu, Bing-Nan and Ma, Yuan-Zhuo and Mei\ss{}ner, Ulf-G. and Noronha-Hostler, Jacquelyn and Plumberg, Christopher and Rodr\'{\i}guez, Tom\'as R. and Roth, Robert and van der Schee, Wilke and Som\`a, Vittorio},
  journal = {Phys. Rev. Lett.},
  volume = {135},
  issue = {1},
  pages = {012302},
  numpages = {9},
  year = {2025},
  month = {Jul},
  publisher = {American Physical Society},
  doi = {10.1103/k8rb-jgvq},
  url = {https://link.aps.org/doi/10.1103/k8rb-jgvq}
}

@article{Giacalone2025b,
  title = {Anisotropic Flow in Fixed-Target $^{208}\mathrm{Pb}+^{20}\mathrm{Ne}$ Collisions as a Probe of Quark-Gluon Plasma},
  author = {Giacalone, Giuliano and Zhao, Wenbin and Bally, Benjamin and Shen, Shihang and Duguet, Thomas and Ebran, Jean-Paul and Elhatisari, Serdar and Frosini, Mikael and L\"ahde, Timo A. and Lee, Dean and Lu, Bing-Nan and Ma, Yuan-Zhuo and Mei\ss{}ner, Ulf-G. and Nijs, Govert and Noronha-Hostler, Jacquelyn and Plumberg, Christopher and Rodr\'{\i}guez, Tom\'as R. and Roth, Robert and van der Schee, Wilke and Schenke, Bj\"orn and Shen, Chun and Som\`a, Vittorio},
  journal = {Phys. Rev. Lett.},
  volume = {134},
  issue = {8},
  pages = {082301},
  numpages = {10},
  year = {2025},
  month = {Feb},
  publisher = {American Physical Society},
  doi = {10.1103/PhysRevLett.134.082301},
  url = {https://link.aps.org/doi/10.1103/PhysRevLett.134.082301}
}

@Article{STAR2024a,
author={Collaboration, S. T. A. R.},
title={Imaging shapes of atomic nuclei in high-energy nuclear collisions},
journal={Nature},
year={2024},
month={Nov},
day={01},
volume={635},
number={8037},
pages={67-72},
issn={1476-4687},
doi={10.1038/s41586-024-08097-2},
url={https://doi.org/10.1038/s41586-024-08097-2}
}

@article{Summerfield2021a,
  title = {$^{16}\mathrm{O}^{16}\mathrm{O}$ collisions at energies available at the BNL Relativistic Heavy Ion Collider and at the CERN Large Hadron Collider comparing $\ensuremath{\alpha}$ clustering versus substructure},
  author = {Summerfield, Nicholas and Lu, Bing-Nan and Plumberg, Christopher and Lee, Dean and Noronha-Hostler, Jacquelyn and Timmins, Anthony},
  journal = {Phys. Rev. C},
  volume = {104},
  issue = {4},
  pages = {L041901},
  numpages = {8},
  year = {2021},
  month = {Oct},
  publisher = {American Physical Society},
  doi = {10.1103/PhysRevC.104.L041901},
  url = {https://link.aps.org/doi/10.1103/PhysRevC.104.L041901}
}

@article{Giacalone2021a,
  title = {Accessing the shape of atomic nuclei with relativistic collisions of isobars},
  author = {Giacalone, Giuliano and Jia, Jiangyong and Som\`a, Vittorio},
  journal = {Phys. Rev. C},
  volume = {104},
  issue = {4},
  pages = {L041903},
  numpages = {6},
  year = {2021},
  month = {Oct},
  publisher = {American Physical Society},
  doi = {10.1103/PhysRevC.104.L041903},
  url = {https://link.aps.org/doi/10.1103/PhysRevC.104.L041903}
}

@article{Zhang2022a,
  title = {Evidence of Quadrupole and Octupole Deformations in $^{96}\mathrm{Zr}+^{96}\mathrm{Zr}$ and $^{96}\mathrm{Ru}+^{96}\mathrm{Ru}$ Collisions at Ultrarelativistic Energies},
  author = {Zhang, Chunjian and Jia, Jiangyong},
  journal = {Phys. Rev. Lett.},
  volume = {128},
  issue = {2},
  pages = {022301},
  numpages = {6},
  year = {2022},
  month = {Jan},
  publisher = {American Physical Society},
  doi = {10.1103/PhysRevLett.128.022301},
  url = {https://link.aps.org/doi/10.1103/PhysRevLett.128.022301}
}

@article{Ryssens2023a,
  title = {Evidence of Hexadecapole Deformation in Uranium-238 at the Relativistic Heavy Ion Collider},
  author = {Ryssens, Wouter and Giacalone, Giuliano and Schenke, Bj\"orn and Shen, Chun},
  journal = {Phys. Rev. Lett.},
  volume = {130},
  issue = {21},
  pages = {212302},
  numpages = {7},
  year = {2023},
  month = {May},
  publisher = {American Physical Society},
  doi = {10.1103/PhysRevLett.130.212302},
  url = {https://link.aps.org/doi/10.1103/PhysRevLett.130.212302}
}

@article{Bally2022a,
  title = {Evidence of the Triaxial Structure of $^{129}\mathrm{Xe}$ at the Large Hadron Collider},
  author = {Bally, Benjamin and Bender, Michael and Giacalone, Giuliano and Som\`a, Vittorio},
  journal = {Phys. Rev. Lett.},
  volume = {128},
  issue = {8},
  pages = {082301},
  numpages = {6},
  year = {2022},
  month = {Feb},
  publisher = {American Physical Society},
  doi = {10.1103/PhysRevLett.128.082301},
  url = {https://link.aps.org/doi/10.1103/PhysRevLett.128.082301}
}

@article{Giacalone2018a,
  title = {Hydrodynamic predictions for 5.44 TeV Xe+Xe collisions},
  author = {Giacalone, Giuliano and Noronha-Hostler, Jacquelyn and Luzum, Matthew and Ollitrault, Jean-Yves},
  journal = {Phys. Rev. C},
  volume = {97},
  issue = {3},
  pages = {034904},
  numpages = {8},
  year = {2018},
  month = {Mar},
  publisher = {American Physical Society},
  doi = {10.1103/PhysRevC.97.034904},
  url = {https://link.aps.org/doi/10.1103/PhysRevC.97.034904}
}

@article{Dobaczewski:2025rdi,
    author = "Dobaczewski, J. and Gade, A. and Godbey, K. and Janssens, R. V. F. and Nazarewicz, W.",
    title = "{Extraction of ground-state nuclear deformations from ultrarelativistic heavy-ion collisions: Nuclear structure physics context}",
    eprint = "2507.05208",
    archivePrefix = "arXiv",
    primaryClass = "nucl-th",
    doi = "10.1103/kngg-1ccb",
    journal = "Phys. Rev. Res.",
    volume = "7",
    number = "4",
    pages = "043159",
    year = "2025"
}

@article{Bofos:2026huw,
    author = "Bofos, Stavros and Bally, Benjamin and Duguet, Thomas and Frosini, Mikael",
    title = "{Imaging two-body correlations in atomic nuclei via low- and high-energy processes}",
    eprint = "2602.09890",
    archivePrefix = "arXiv",
    primaryClass = "nucl-th",
    month = "2",
    year = "2026"
}

@article{Frosini21a,
	author = {{Frosini, M.} and {Duguet, T.} and {Bally, B.} and {Beaujeault-Taudi\`ere, Y.} and {Ebran, J.-P.} and {Som\`a, V.}},
	title = {In-medium k-body reduction of n-body operators - A flexible symmetry-conserving approach based on the sole one-body density matrix},
	DOI= "10.1140/epja/s10050-021-00458-z",
	url= "https://doi.org/10.1140/epja/s10050-021-00458-z",
	journal = {Eur. Phys. J. A},
	year = 2021,
	volume = 57,
	number = 4,
	pages = "151",
}

@article{Hebeler11a,
  title = {Improved nuclear matter calculations from chiral low-momentum interactions},
  author = {Hebeler, K. and Bogner, S. K. and Furnstahl, R. J. and Nogga, A. and Schwenk, A.},
  journal = {Phys. Rev. C},
  volume = {83},
  issue = {3},
  pages = {031301},
  numpages = {5},
  year = {2011},
  month = {Mar},
  publisher = {American Physical Society},
  doi = {10.1103/PhysRevC.83.031301},
  url = {https://link.aps.org/doi/10.1103/PhysRevC.83.031301}
}

@article{Burvenich:2002,
    author = "Burvenich, T. and Madland, D. G. and Maruhn, J. A. and Reinhard, P. G.",
    title = "{Nuclear ground state observables and QCD scaling in a refined relativistic point coupling model}",
    eprint = "nucl-th/0111012",
    archivePrefix = "arXiv",
    doi = "10.1103/PhysRevC.65.044308",
    journal = "Phys. Rev. C",
    volume = "65",
    pages = "044308",
    year = "2002"
}

@article{Niemi:2016,
  title = {Event-by-event fluctuations in a perturbative QCD + saturation + hydrodynamics model: Determining QCD matter shear viscosity in ultrarelativistic heavy-ion collisions},
  author = {Niemi, H. and Eskola, K. J. and Paatelainen, R.},
  journal = {Phys. Rev. C},
  volume = {93},
  issue = {2},
  pages = {024907},
  numpages = {29},
  year = {2016},
  month = {Feb},
  publisher = {American Physical Society},
  doi = {10.1103/PhysRevC.93.024907},
  url = {https://link.aps.org/doi/10.1103/PhysRevC.93.024907}
}

@article{Yao:2014,
    author = "Yao, J. M. and Hagino, K. and Li, Z. P. and Meng, J. and Ring, P.",
    title = "{Microscopic benchmark study of triaxiality in low-lying states of 76Kr}",
    eprint = "1403.4812",
    archivePrefix = "arXiv",
    primaryClass = "nucl-th",
    doi = "10.1103/PhysRevC.89.054306",
    journal = "Phys. Rev. C",
    volume = "89",
    number = "5",
    pages = "054306",
    year = "2014"
}

@article{Balian:1969,
    author = "Balian, R. and Brezin, E.",
    title = "{Nonunitary bogoliubov transformations and extension of wick's theorem}",
    doi = "10.1007/BF02710281",
    journal = "Nuovo Cim. B",
    volume = "64",
    pages = "37--55",
    year = "1969"
}

@article{Ke:2025,
    author = "Ke, Weiyao",
    title = "{From wave-function to fireball geometry: the role of a restored broken symmetry in ultra-relativistic collisions of deformed nuclei}",
    eprint = "2509.09549",
    archivePrefix = "arXiv",
    primaryClass = "nucl-th",
    month = "9",
    year = "2025"
}

\end{document}